# Advanced LIGO

LIGO Scientific Collaboration (August 2014 LSC author list)

Email: pf@ligo.mit.edu

**Abstract**

The Advanced LIGO gravitational wave detectors are second generation instruments designed and built for the two LIGO observatories in Hanford, WA and Livingston, LA. The two instruments are identical in design, and are specialized versions of a Michelson interferometer with 4 km long arms. As in initial LIGO, Fabry-Perot cavities are used in the arms to increase the interaction time with a gravitational wave, and power recycling is used to increase the effective laser power. Signal recycling has been added in Advanced LIGO to improve the frequency response. In the most sensitive frequency region around 100 Hz, the design strain sensitivity is a factor of 10 better than initial LIGO. In addition, the low frequency end of the sensitivity band is moved from 40 Hz down to 10 Hz. All interferometer components have been replaced with improved technologies to achieve this sensitivity gain. Much better seismic isolation and test mass suspensions are responsible for the gains at lower frequencies. Higher laser power, larger test masses and improved mirror coatings lead to the improved sensitivity at mid- and high-frequencies. Data collecting runs with these new instruments are planned to begin in mid-2015.

## 1. Introduction

From its inception, the LIGO project was planned to involve the development and operation of multiple generations of increasingly sensitive gravitational wave detectors. The Advanced LIGO detectors are the second generation of interferometers designed and built for the two observatories operated by the LIGO Laboratory: one on the Hanford site in Washington state, and the other in Livingston Parish, Louisiana. The Initial LIGO detectors, constructed in the late 1990s, operated at design sensitivity in a continuous data-taking mode from November 2005 to September 2007 [1]. Subsequently, upgrades were made to the interferometer laser sources and readout systems [2], which improved the strain sensitivity at the most sensitive frequencies by approximately 30%; the instrument root-mean-square (rms) strain noise reached an unprecedented level of $2\times10^{-22}$ in a 100 Hz band. In this configuration, Enhanced LIGO collected data from July 2009 to October 2010. Although no gravitational wave signals were detected with Initial or Enhanced LIGO, they produced several interesting astrophysical results; representative examples are listed in reference [3].

Compared to Initial LIGO, Advanced LIGO is designed to provide a factor of 10 increase in strain sensitivity over a broad frequency band, and to extend the low end of the band to 10 Hz (from 40 Hz). As the probed volume of the universe scales as the cube of the strain sensitivity, this represents an enormous increase (of order $10^3$x) in the number of potential astrophysical sources detectable by these instruments. At design sensitivity, Advanced LIGO is likely to detect dozens of compact binary coalescence sources per year [4].

Housed within the vacuum system and facilities built for Initial LIGO, Advanced LIGO completely replaces the interferometer components with new designs and technology. Each observatory hosts one Advanced LIGO interferometer with 4km long arms. A third interferometer (originally planned as a second detector at Hanford) is consigned for a planned later installation at a new, currently undetermined site in India. All three interferometers are intended to be identical in design and expected performance. The addition of the third site (India) will provide significantly better source localization for the Advanced LIGO network [5].

US National Science Foundation funding for the construction and installation of Advanced LIGO began in April 2008. Installation of the new hardware began in early 2011, and was completed for the Livingston detector (L1) in mid-2014; installation of the Hanford detector (H1) will be completed by the end of 2014.

## 2. Interferometer configuration and system design

The optical configuration of the Advanced LIGO interferometer is shown in figure 1. The basis of the design is a Michelson interferometer with a Fabry-Perot resonant cavity in each arm to build up the phase shift produced by an arm length change. Power recycling is another standard feature of such interferometers: the power recycling mirror (PRM) forms a resonant cavity between the laser source and the Michelson to increase the effective laser power.

With Advanced LIGO, signal recycling [6] has been added to the interferometer. The signal recycling mirror (SRM) at the anti-symmetric output of the Michelson is used to effectively lower the arm cavity finesse for gravitational wave signals and thereby maintain a broad detector frequency response. The choice of SRM transmission and tuning is discussed in the section on sensitivity (3). In principle, signal recycling can also be used to create a narrowband mode of operation, with enhanced sensitivity at, e.g., a likely pulsar frequency, though with higher noise at other frequencies. Though there are no current plans for narrowband operation, this option may be taken in the future.

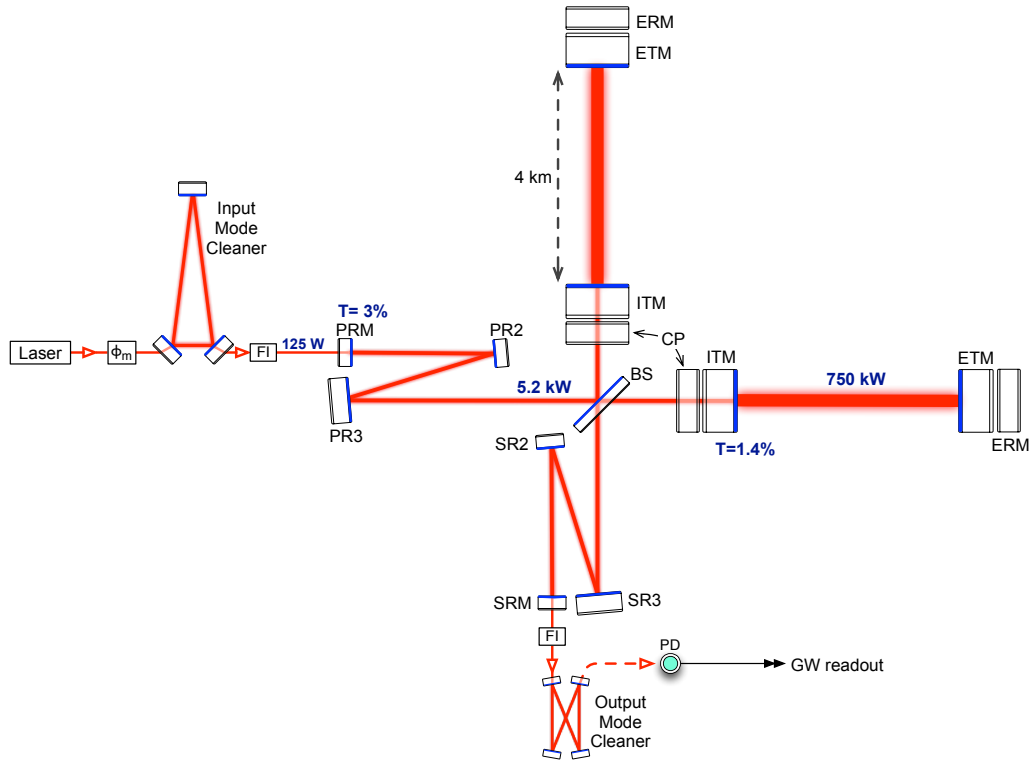

**Figure 1.** Advanced LIGO optical configuration. ITM: input test mass; ETM: end test mass; ERM: end reaction mass; CP: compensation plate; PRM: power recycling mirror; PR2/PR3: power recycling mirror 2/3; BS: 50/50 beam splitter; SRM: signal recycling mirror; SR2/SR3: signal recycling mirror 2/3; FI: Faraday isolator; $\varphi_m$: phase modulator; PD: photodetector. The laser power numbers correspond to full-power operation. All of the components shown, except the laser and phase modulator, are mounted in the LIGO ultra-high vacuum system on seismically isolated platforms.

The top-level parameters of the interferometers are listed in Table 1. The motivations behind these and other system design choices are described in this section. The various interferometer subsystems and components are described in section 4.

**Table 1.** Main parameters of the Advanced LIGO interferometers. PRC: power recycling cavity; SRC: signal recycling cavity.

| Parameter | Value |
|---|---|
| Arm cavity length | 3994.5 m |
| Arm cavity finesse | 450 |
| Laser type and wavelength | Nd:YAG, $\lambda = 1064$ nm |
| Input power, at PRM | up to 125 W |
| Beam polarization | linear, horizontal |
| Test mass material | Fused silica |
| Test mass size & mass | 34cm diam. x 20cm, 40 kg |
| Beam radius ($1/e^2$), ITM / ETM | 5.3 cm / 6.2 cm |
| Radius of curvature, ITM / ETM | 1934 m / 2245 m |
| Input mode cleaner length & finesse | 32.9 m (round trip), 500 |
| Recycling cavity lengths, PRC / SRC | 57.6 m / 56.0 m |

*2.1 Arm cavity design*

With signal recycling, the quantum noise-limited strain sensitivity is in principle independent of the arm cavity finesse, and depends only on the power stored in the arms; thus the choice of finesse is guided by other practical considerations. Higher arm cavity finesse carries the benefits of lower laser power in the power recycling cavity and reduced coupling from the vertex Michelson degree-of-freedom. Lower finesse reduces the sensitivity to optical loss and noise in the signal recycling cavity. The arm cavity finesse of 450 represents a trade-off between these effects.

In order to reduce test mass thermal noise, the beam size on the test masses is made as large as practical so that it averages over more of the mirror surface. The dominant noise mechanism here is mechanical loss in the dielectric mirror coatings, for which the displacement thermal noise scales inversely with beam size. This thermal noise reduction is balanced against increased aperture loss and decreased mode stability with larger beams. The slightly asymmetric design of the arm cavity takes advantage of the fact that the ITMs contribute less to thermal noise because their coatings are half as thick as on the ETMs. Therefore the beam can be somewhat smaller on the ITMs—with negligible increase in thermal noise—in order to limit aperture losses in the beam splitter and recycling cavities; the ETM beam size is maximized to reduce this thermal noise contribution. The resonator *g* parameter product for the arm cavity is $g_1 g_2 = 0.83$, which is approaching the stability limit of $g_1 g_2 \rightarrow 1$. Lower thermal noise thus comes at the expense of greater sensitivity to angular misalignment.

The specified mirror beam sizes can be achieved with either of two designs: a nearly-planar cavity or a nearly-concentric one. The nearly-concentric design is preferred for its higher stability with high stored power operation when torques due to optical radiation pressure become significant. In this case, the torsional mode with the higher optical stiffness is statically stable, whereas it would be statically unstable for a near-planar design [7].

*2.2 Recycling cavity design*

Both the power recycling and signal recycling cavities are designed to be stable for the fundamental Gaussian modes they support. That is, the fundamental cavity mode accrues a non-negligible Gouy phase in a one-way propagation through the cavity. The benefit of this design (new in Advanced LIGO) is that both recycling cavities have well defined spatial eigenmodes and transversal mode-spacings much greater than the linewidth of the cavities [8]. The modes become less sensitive to mirror imperfections, resulting in more efficient signal detection. The stable design results in the folded layout shown in figure 1 for these cavities. Each of the six recycling cavity mirrors is a curved optic; they produce a one-way Gouy phase of 25 and 19 degrees in the power and signal recycling cavities, respectively, and transform the beam radius from 5.3 cm at the ITMs to 2.2 mm at the PRM, and 2.1 mm at the SRM.

*2.3 Gravitational wave readout*

Readout of the gravitational wave signal is accomplished using an output mode cleaner in conjunction with homodyne, or DC detection. In this scheme, a local oscillator field is generated by offsetting the arm cavities slightly from their resonance (typically a few picometers), thereby pulling the Michelson slightly off the dark fringe [2]. The output mode cleaner filters out non-$TEM_{00}$ mode carrier power, and any power in RF modulation sidebands, so that only the carrier $TEM_{00}$ mode is detected. This greatly reduces the power that must be detected at the output. Homodyne readout is a significant departure from the heterodyne readout used in Initial LIGO. Compared to heterodyne detection, it is less susceptible to a number of technical noise couplings, but its primary benefits lie in lower quantum noise [9] and compatibility with the future use of squeezed light [10].

*2.4 High power effects*

Several effects may hinder interferometer operation at high power and need to be considered in the design: optical distortion produced by thermal effects from absorbed power; optical torques that can significantly impact alignment stability; and parametric instabilities arising from coupling between test mass acoustic modes and higher-order optical modes of the arm cavities.

   As mentioned above, the torque induced by radiation pressure becomes comparable to the mechanical restoring torque of the test mass suspension, and must be accounted for in the angular controls system. This problem is addressed in reference [11], where it is shown that the unstable alignment modes have very low frequency, and can be readily stabilized with a suitable control filter.

   The dominant source of thermal distortion is thermal lensing in the ITM substrates due to power absorbed in the ITM reflective coatings. Next in importance, at high power, coating absorption in the ITMs and ETMs causes a non-negligible change their radii of curvature through thermo-elastic distortion. The thermal compensation system, described in section 4.6, is designed to compensate both of these effects. The compensation plates (CP) shown in figure 1 allow some of the distortion correction to be applied to these elements, which are more noise-tolerant than the ITMs. The use of ultra-low absorption fused silica for the ITMs, the BS, and the CPs ensures that power absorbed in the bulk is negligible compared to that absorbed in the test mass high reflectivity coatings.

Parametric acousto-optic couplings have the potential to lead to unstable build-up of such coupled, higher-order modes [12]. Unchecked, this could impose a limit on the power stored in the arms. Uncertainties in the parameters relevant to the process prevent an exact calculation of the situation. Instead, statistical analyses can indicate the probability of instability at a given power level. The general conclusion of these analyses is that there may be 5-10 modes per test mass that could become unstable at full power. One or more of several mitigation methods may thus need to be applied. The simplest is to use the thermal compensation actuators to slightly change the radius of curvature of one or more test masses; this would shift the eigen-frequency of the higher-order optical mode to avoid overlap with the corresponding acoustic mode. A second technique would be to actively damp any unstable acoustic modes. Each test mass is equipped with an electro-static actuator (see section 4.4.2) that can be used to apply a damping feedback force at the acoustic mode frequencies. A third technique, still in the research phase, is to apply passive tuned dampers to the test masses.

## 3. Strain sensitivity

The interferometer noise floor is determined by the fundamental noise sources, quantum noise and thermal noise. Thermal noise is determined by fixed parameters in the interferometer, such as material properties and beam size. Quantum noise, on the other hand, depends on the readily variable input laser power, and the (less readily) changeable SRM transmission. Other noise sources, such as laser frequency or amplitude noise, photodetector dark noise, actuator noise, etc., are classified as 'technical' noises. Technical noises are controlled by design so that the equivalent strain noise of each source is no greater than 10% of the target strain sensitivity throughout the detection band (10—7000 Hz). As these noise sources are typically statistically independent, they add as a root-square-sum to the total; an individual 10% noise source thus increases the noise floor by only 0.5%.

The projected strain noise spectrum for the nominal Advanced LIGO mode of operation is shown in figure 2. In the nominal mode, the input power at the PRM is 125 W, the SRM transmission is 20%, and the signal recycling cavity has zero detuning. The individual noise terms are described in the following sections.

Beyond the strain noise spectrum, a standard figure of merit for detector sensitivity is the distance to which the GW signal emitted by a binary neutron star (BNS) coalescence is detectable. The BNS *range* is defined as the volume- and orientation-averaged distance at which a coalescence gives a matched filter signal-to-noise ratio of 8 in a single detector [13]. The BNS range for the strain noise curve in figure 2 is 190 Mpc.

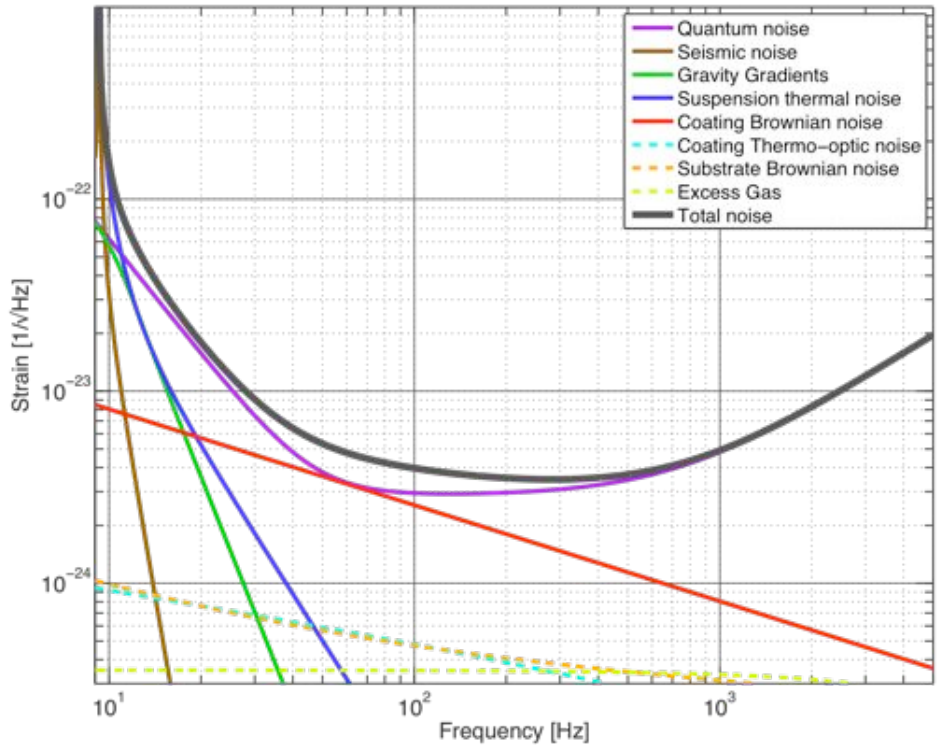

**Figure 2** Principal noise terms for the nominal (high power, broadband) mode of operation of Advanced LIGO.

*3.1 Quantum noise*
Quantum noise encompasses the effects of statistical fluctuations in detected photon arrival rate (shot noise) and radiation pressure due to photon number fluctuations. Quantum noise is calculated using the formulation of Buonanno and Chen [14]. We assume 75 ppm round-trip loss in each arm cavity, and $10^{-3}$ loss in the power recycling cavity, which leads to 5.2 kW of power at the beamsplitter and 750 kW of power in each arm cavity. A detection efficiency of 90% is assumed; this accounts for finite transmission through the output Faraday isolator and output mode cleaner, as well as photodetector quantum efficiency.

*3.2 Test mass thermal noise*
Coating Brownian noise is the dominant of the various test mass thermal noise terms. It arises from mechanical dissipation in the coatings, and is calculated according to reference [15]. The coating design and material parameters are described in section 4.3. Coating thermo-optic noise arises from thermal dissipation in the coatings, producing noise via the thermoelastic and thermorefractive coefficients of the coating materials. The two effects are calculated coherently, according to reference [16]. Mechanical loss in the bulk fused silica is responsible for the substrate Brownian noise term. Reference [17] provides the calculation for this term, using the bulk and surface loss model for fused silica found in reference [18].

*3.3 Suspension thermal noise*
Thermal noise in the test mass suspension is primarily due to loss in the fused silica fibres used in the final suspension stage. As described in the suspension design section (4.4.2), these four glass fibres have a circular, but variable diameter cross-section: they are thin in the main (middle) section of the fibre, and about twice as thick near the ends. This geometry minimizes thermal noise, while keeping the fibre violin mode frequency high (510 Hz fundamental) and the vertical stretching mode frequency low (9Hz). The thermal noise is calculated using a finite-element model of the last suspension stage, including loss terms for the bulk, the surface, and the thermoelastic components of the fibre material [19].

*3.4 Gravity gradients*
Seismic waves produce density perturbations in the earth close to the test masses, which in turn produce fluctuating gravitational forces on the masses. This seismic gravity-gradient noise is estimated using the transfer function formulation of reference [20], and a representative model for the seismic motion at an observatory site. The latter can be quite variable over time, and gravity gradient noise is predicted to be several times higher than shown in figure 2 for some time periods [21].

Figure 2 also shows the strain noise from the transmission of seismic noise through the seismic isolation and suspension systems to the test masses (the 'Seismic noise' curve). Due to the large mechanical isolation, this noise is negligible above 11 Hz.

*3.5 Residual gas noise*
Residual gas in the 4 km long beam tubes will lead to statistical variations in the column density of gas particles in the beam path, producing fluctuations in the effective refractive index along the path. The resulting optical path length noise is modeled by calculating the impulsive change in the cavity field's phase as a molecule moves through the beam, and integrating over the molecular velocity distribution [22]. The noise curve includes only the most dominant residual gas component, hydrogen, at a pressure of $4\times10^{-7}$ pascals.

Though not included in the figure 2 noise curve, residual gas in the test mass vacuum chambers will contribute some damping to the test mass suspension, potentially increasing the suspension thermal noise. This damping effect is increased by the relatively narrow gap between the test mass and its suspended reaction mass — so-called thin-film damping [23]. Gas damping noise is most significant in the 10—40 Hz band, where it falls very nearly as $f^{-2}$. At an anticipated chamber pressure of $7\times10^{-7}$ pascals of $H_2$, the resulting strain noise is $5\times10^{-24}/\sqrt{Hz}$ at 20 Hz – a factor of 3-4 below the interferometer strain noise in figure 2. To mitigate the effect, this gap is larger for the ITMs (20 mm) than the ETMs (5 mm), since the former do not require as much electro-static actuation force. This noise term is not included in figure 2 because it will eventually be made negligible (i.e., another factor of 3 reduction), through some combination of lower chamber pressure, an increased gap for the ETMs, and possibly a more complicated (ring-like) geometry for the end reaction masses.

*3.6 Other modes of operation*

As mentioned above, the laser input power and the signal recycling cavity parameters can be varied to implement different modes of operation than the nominal mode represented by figure 2. We illustrate the potential parameter space with two particular alternate interferometer modes: a mode optimized for low input power; a mode optimized for BNS detection. The low power mode is of interest because achieving operation at full power is likely to take extended commissioning time; early operations and observation runs will therefore be carried out at reduced power. The BNS optimized mode shows the potential sensitivity to this particular source, and the corresponding trade-off in broadband sensitivity. Table 2 gives the parameters for the two modes, and figure 3 shows their strain noise spectra.

**Table 2** Interferometer parameters for two alternate modes of operation

| Mode | Input power | SRM transmission | SRC detuning | BNS range |
|---|---|---|---|---|
| Low power | 25 W | 35% | 0 | 160 Mpc |
| BNS optimized | 125 W | 20% | 16 deg. | 210 Mpc |

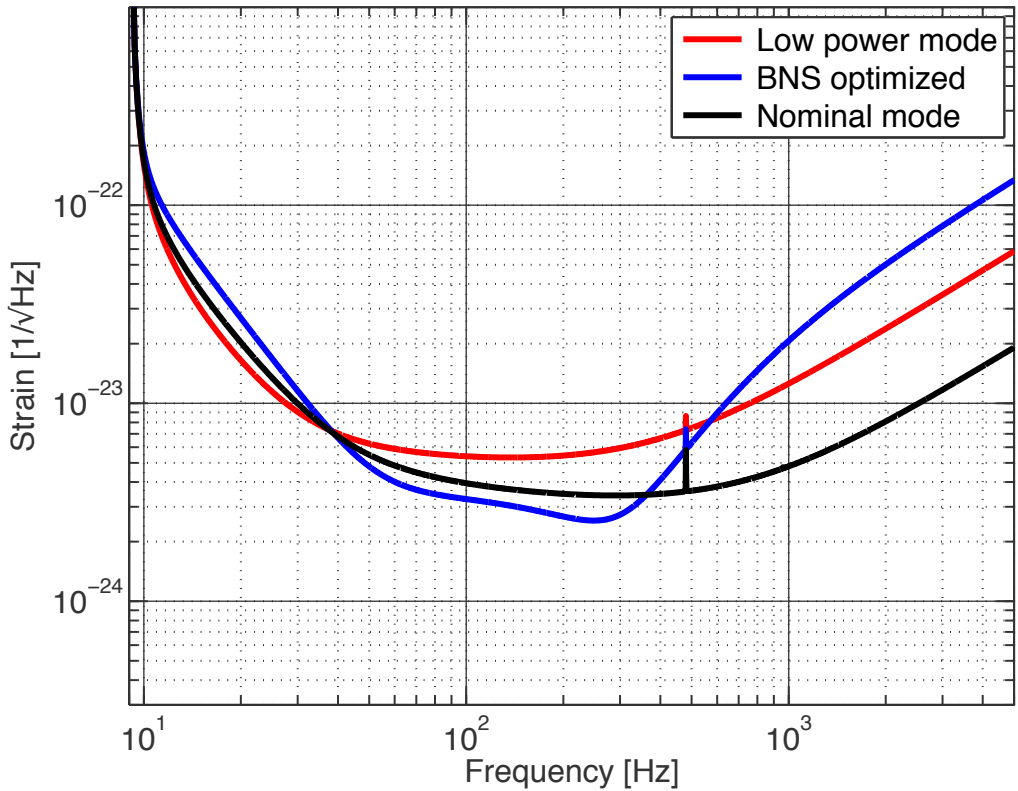

**Figure 3** Advanced LIGO strain noise spectrum for the modes defined in table 2, and for the nominal sensitivity shown in figure 2. The feature at 500 Hz is the (unresolved) fundamental vibrational mode of the test mass suspension fibres.

## 4. Detector subsystems

*4.1 Laser source*

The interferometer employs a multi-stage Nd:YAG laser that can supply up to 180 W at the laser system output. The pre-stabilized laser (PSL) system [24] consists of this laser light source, and control systems that stabilize the laser in frequency, beam direction, and intensity. The laser was developed and supplied by the Max Planck Albert Einstein Institute in collaboration with Laser Zentrum Hannover e.V. A schematic drawing of the PSL is shown in Figure 4.

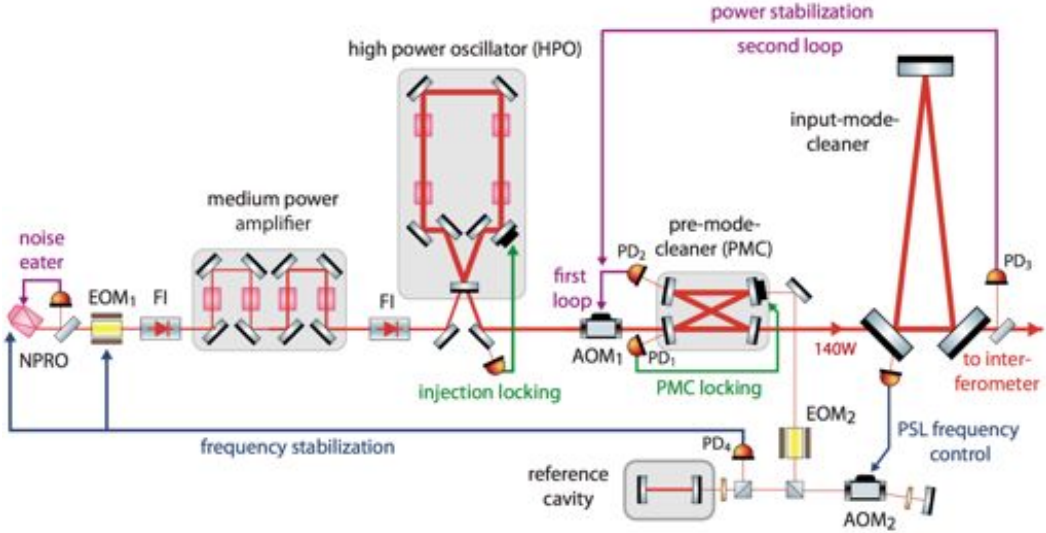

**Figure 4** Schematic of the pre-stabilized laser system. The Input Mode Cleaner is discussed in section 4.2. AOM: acousto-optic modulator; EOM: electro-optic modulator; FI: Faraday isolator; PD: photodetector.

The laser comprises three stages. The first stage is a commercial non-planar ring-oscillator (NPRO) manufactured by InnoLight GmbH. The second stage (medium power amplifier) is a single-pass amplifier that boosts the NPRO power to 35 W, and the third stage is an injection-locked ring oscillator with a maximum output power of about 220 W. All stages are pumped by laser diodes; the pump diodes for the second and third stages are fibre coupled and located far from the laser table for easier management of power and cooling. The system may be configured for 35 W output by using the NPRO and medium power stages, bypassing the high power oscillator; this configuration will be in the early operations of Advanced LIGO.

The source laser is passed through the pre-mode cleaner (PMC). The PMC is a bow-tie cavity (2m round trip length) designed to strip higher-order modes from the beam, to reduce beam jitter (amplitude reduction factor for $TEM_{01}/TEM_{10}$ modes of 63), and to provide low-pass filtering for RF intensity fluctuations (cavity pole at 560 kHz).

Intensity stabilization requirements are quite stringent across the GW band. At full laser power, radiation pressure effects in the arm cavities lead to a residual intensity noise specification of $2 \times 10^{-9}$ $Hz^{-1/2}$ at 10 Hz, at the input to the PRM. This stability is achieved through cascaded sensors and feedback loops. The first sensor measures the power in the PMC mode, and the second sensor samples light transmitted through the Input Mode Cleaner. To ensure that an accurate low-noise measurement is made, the second sensor profits from the beam stability post-mode-cleaner, is in vacuum, and consists of multiple diodes used at low power. An acousto-optic modulator ($AOM_1$) is the actuator for the control loop.

The initial frequency stabilization of the laser is performed in the PSL by locking its frequency to an isolated, high-finesse reference cavity (bandwidth of 77 kHz) using the standard reflection locking technique. Three actuators are used to provide wide bandwidth and large dynamic range: a PZT attached to the NPRO crystal; an electro-optic modulator ($EOM_1$)—used as a broadband, phase corrector—between the NPRO and medium power amplifier; and the NPRO crystal temperature for slow, wide-range control. The servo bandwidth is 400 kHz. The beam used for this frequency pre-stabilization is taken after the PMC, and is double-passed through an AOM before being sent to the reference cavity. This AOM is driven by a voltage-controlled oscillator to provide laser frequency actuation (1 MHz range, 100 kHz bandwidth) for subsequent stages of stabilization.

Figure 5 shows a mode scan of the L1 high-power laser beam when operating at full output power. The total power in higher-order (non-$TEM_{00}$) modes is 5.3% of the total. This is measured before the PMC, which will filter out this high-order mode content.

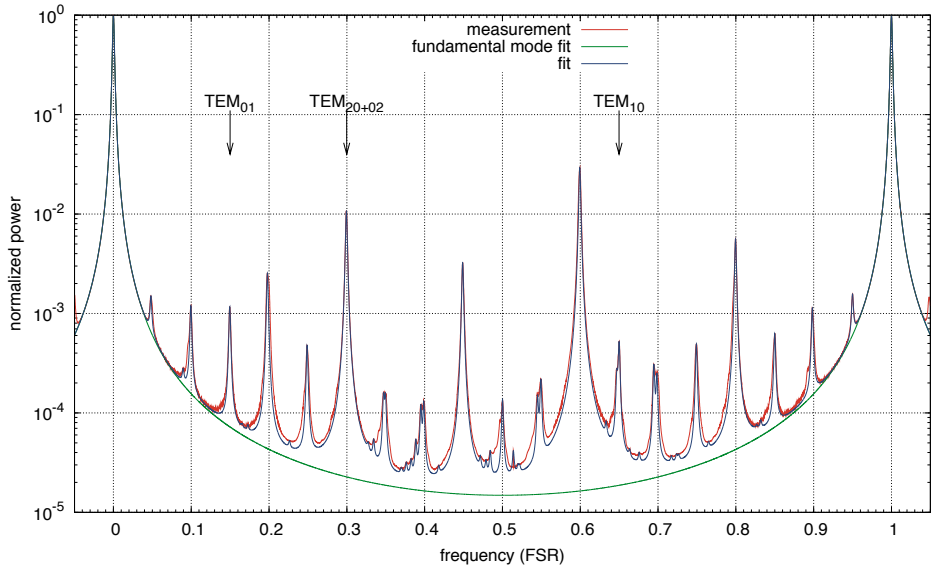

**Figure 5** Mode scan of the high power laser used on the L1 interferometer. Close to 95% of the power is in the $TEM_{00}$ mode.

*4.2 Input optics*

The input optics subsystem (IO) accepts the light from the pre-stabilized laser, stabilizes and conditions it, and performs matching and pointing into the main interferometer core optics. A schematic view is shown in Figure 6. The optical efficiency of the IO, from the PSL output to the PRM, is designed to be at least 75%. The system can be broken down into the following functional units.

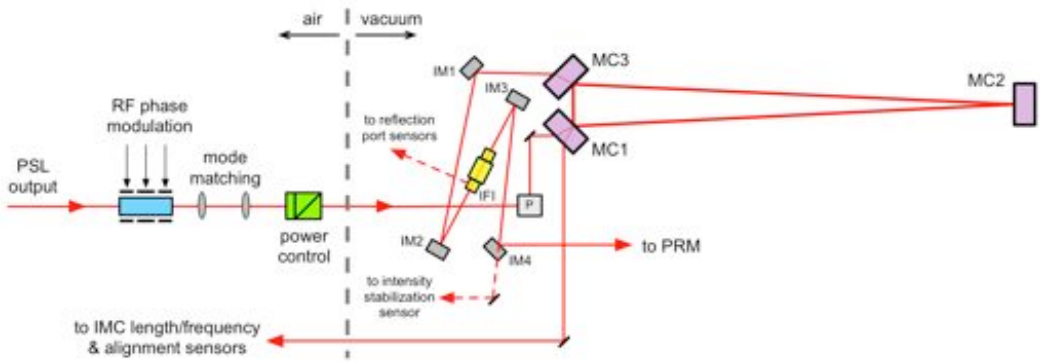

**Figure 6** Schematic of the input optics components. IFI: input Faraday isolator. P: periscope. IMn: input mirror 1-4.

*4.2.1 RF Modulation*. Radio-frequency (RF) phase modulation is impressed on the beam and used for global sensing of the interferometer (see section 4.8) and for sensing of the IMC. Three modulation frequencies are applied, all with small modulation depth: 9 MHz and 45 MHz for the main interferometer sensing; 24 MHz for IMC sensing. The modulator must exhibit good phase modulation efficiency, low residual amplitude modulation, and negligible thermal lensing up to a beam power of 165 W [25]. The EOM uses a 40mm long, 4×4 mm cross-section, wedged RTP (rubidium titanyl phosphate) crystal as the electro-optic material. Three pairs of electrodes are used to apply the three modulation frequencies to the crystal.

*4.2.2 Input Mode Cleaner*. The IMC supplies a key function of the IO system: to stabilize the PSL beam in position and mode content, and to provide a high-quality laser frequency reference. The beam pointing stability at the input to the PRM, expressed relative to the beam radius and beam divergence angle, must be $<1\times10^{-8}$ $Hz^{-1/2}$ at 100 Hz. The frequency stability at the IMC output must be $<1\times10^{-3}$ $Hz/Hz^{1/2}$ at 100 Hz.

The IMC is a three-mirror ring cavity used in transmission. Each mirror is suspended in a triple pendulum suspension with metal wire loops (see section 4.4) to provide vibration isolation and acceptable thermal noise. The IMC finesse is 500, the round trip length is 32.9 m, and the linewidth is 18 kHz. The 24 MHz phase modulation sidebands are used both for reflection locking and wavefront-sensor alignment control of the IMC. Additionally, the IMC is used as a frequency actuator for the final level of frequency stabilization to the long arm cavities.

*4.2.3 Faraday Isolator*. An in-vacuum Faraday Isolator is mounted between the IMC output and the PRM; the isolator extracts the beam reflected from the interferometer and prevents this beam from creating parasitic interference in the input chain. The isolator must deliver a minimum of 30dB of isolation up to 125 W of laser power. To compensate thermally induced birefringence, the isolator uses two terbium gallium garnet (TGG) crystals for Faraday rotation, with a quartz rotator in between. Compensation of thermal lensing is achieved by incorporating a negative *dn/dT* material (deuterated potassium dihydrogen phosphate, or 'DKDP') in the assembly [26].

*4.2.4 Mode Matching*. The IMC output beam must be matched to the interferometer mode, with a targeted efficiency of 95% or better. This mode matching is accomplished with two curved mirrors, in combination with two flat mirrors used for beam routing. All four of these reflective optics are mounted in single stage suspensions for vibration isolation; the suspensions also have actuators to provide remote steering capability (see section 4.4).

*4.3 Core optics*

The 'Core Optics' are central to the interferometer performance. For each interferometer they include (see Table 3):

- Two input and two end test masses which form the Fabry-Perot arms
- A 50/50 beamsplitter at the vertex of the Michelson interferometer
- Two compensation plates that serve as actuation reaction masses for the input test masses, and to which thermal compensation can be applied
- Two actuation reaction masses for the end test masses
- Four reflective curved mirrors in the signal and power recycling cavities
- Signal and power recycling mirrors

**Table 3** Parameters of the core optics. ETM/ITM: end/input test mass; CP: compensation plate; ERM: end reaction mass; BS: beam splitter; PR3/2: power recycling mirror 3/2; SR3/2: signal recycling mirror 3/2; PRM/SRM: power/signal recycling mirror. ROC: radius of curvature; AR: anti-reflection. Transmission values are at 1064 nm, except for those in parentheses, which are for 532 nm.

| Optic | Dimensions: diam.×thickness | Mass | Transmission | ROC | Beam size ($1/e^2$ radius) |
|---|---|---|---|---|---|
| ITM | 34×20 cm | 40 kg | 1.4% (0.5-2%) | 1934 m | 5.3 cm |
| ETM | 34×20 cm | 40 kg | 5 ppm (1-4%) | 2245 m | 6.2 cm |
| CP | 34×10 cm | 20 kg | AR< 50 ppm | flat | 5.3 cm |
| ERM | 34×13 cm | 26 kg | AR< 1000 ppm | flat | 6.2 cm |
| BS | 37×6 cm | 14 kg | 50% | flat | 5.3 cm |
| PR3 | 26.5×10 cm | 12 kg | < 15 ppm | 36.0 m | 5.4 cm |
| SR3 | 26.5×10 cm | 12 kg | < 15 ppm | 36.0 m | 5.4 cm |
| PR2 | 15×7.5 cm | 2.9 kg | 225 ppm (>90%) | -4.56 m | 6.2 mm |
| SR2 | 15×7.5 cm | 2.9 kg | < 15 ppm | -6.43 m | 8.2 mm |
| PRM | 15×7.5 cm | 2.9 kg | 3.0% | -11.0 m | 2.2 mm |
| SRM | 15×7.5 cm | 2.9 kg | 20% | -5.69 m | 2.1 mm |

All the core optics are made with fused silica substrates. Their fabrication involves three components: material type; substrate polishing; and coatings. The test masses naturally have the most stringent requirements for all three aspects.

The fused silica material used for the input test masses, the compensation plates, and the beamsplitter is Heraeus Suprasil 3001. This is an ultra-low absorption grade, with absorption at 1064 nm of < 0.2 ppm/cm. The material also has a low level of inhomogeneity and low mechanical loss. The material for the other core optics is less critical, and less expensive grades of fused silica are used (ETMs use Heraeus Suprasil 312).

The surface quality required of the polishing is determined largely by the targeted round-trip optical loss in the arm cavities. For a given input laser power, this loss determines the achievable stored power in the arms, and thus the quantum noise level of the interferometer. For Advanced LIGO the arm round-trip loss goal is set at 75 ppm. This would allow a total arm cavity stored power (both arms) of up to: 1/75 ppm = $1.3 \times 10^4$ times the input power. Table 4 lists the test mass substrate polishing requirements, as well as typical achieved levels. The substrates for the large core optics are produced by a combination of super-polishing for small scale smoothness, followed by ion-beam milling to achieve the large scale uniformity.

**Table 4** Polishing specifications and results for the test masses

|  | Surface error, central 160 mm diam., power & astigmatism removed, rms | | Radius of curvature spread |
|---|---|---|---|
|  | > 1 mm$^{-1}$ | 1—750 mm$^{-1}$ |  |
| Specification | < 0.3 nm | < 0.16 nm | -5, +10 m |
| Actual | 0.08—0.23 nm | 0.07—0.14 nm | -1.5, +1 m |

All the optical coatings are ion-beam sputtered, multi-layer dielectrics. The test masses are coated by Laboratoire Matériaux Avancés (LMA, Lyon, France); all other large core optics are coated by CSIRO (Sydney, Australia). The critical properties of the coating materials are low optical absorption, low scatter, and low mechanical loss (particularly for the test masses). For all optics other than the test masses, the coatings are traditional alternating layers of silicon-dioxide and tantalum pentoxide, each of ¼-wavelength optical thickness. For the test masses, the $TaO_5$ is doped with 25% titanium dioxide, a recipe that reduces the mechanical loss by about 40% [28]. In addition, the layer thicknesses are altered: the $SiO_2$ layers are a little thicker and the Ti-$TaO_5$ layers are a little thinner than a ¼-wavelength. This design not only achieves the transmission specifications at 532 nm, but it also gives a further, modest reduction in thermal noise because less Ti-$TaO_5$ is used; see Table 5 for the coating material parameters that are used to calculate the thermal noise curve shown in Figure 2.

**Table 5** Test mass coating material parameters used to calculate coating thermal noise (i.e., this represents our model of the coating).

| Parameter | Low-index: silica | High-index: tantala |
|---|---|---|
| Mechanical loss | $4\times10^{-5}$ [29] | $2.3\times10^{-4}$ [30] |
| Index of refraction | 1.45 | 2.0654 |
| $dn/dT$ | $8\times10^{-6}$ / K | $1.4\times10^{-5}$ / K [30] |
| Thermal expansion coefficient | $5.1\times10^{-7}$ / K | $3.6\times10^{-6}$ / K [30] |
| Young's modulus | 72 GPa [29] | 140 GPa [30] |
| Layer optical thickness, ITM / ETM | $0.308\,\lambda$ / $0.27\,\lambda$ | $0.192\,\lambda$ / $0.23\,\lambda$ |

All core optics are characterized with high precision metrology, before and after coating. As an example, Figure 7 shows power spectra of the measured phase maps of the test masses used in one arm of the L1 interferometer. The coating contributes additional non-uniformity greater than the substrate at larger spatial scales. Comparing the phase map residuals over the central 160 mm diameter, after subtracting tilt and power, gives:

- **ETM** substrate: 0.18 nm-rms    coated: 0.69 nm-rms
- **ITM** substrate: 0.15 nm-rms    coated: 0.31 nm-rms

Optical simulations of an arm cavity formed with these two test mass mirrors predict a total round trip loss of about 44 ppm (20 ppm due to distortions captured in the phasemap, 20 ppm due to smaller scale roughness that causes wider angle scatter, 4 ppm from ETM transmission). This is well within the goal of 75 ppm, though particulate contamination of installed optics can contribute a few tens of ppm additional loss.

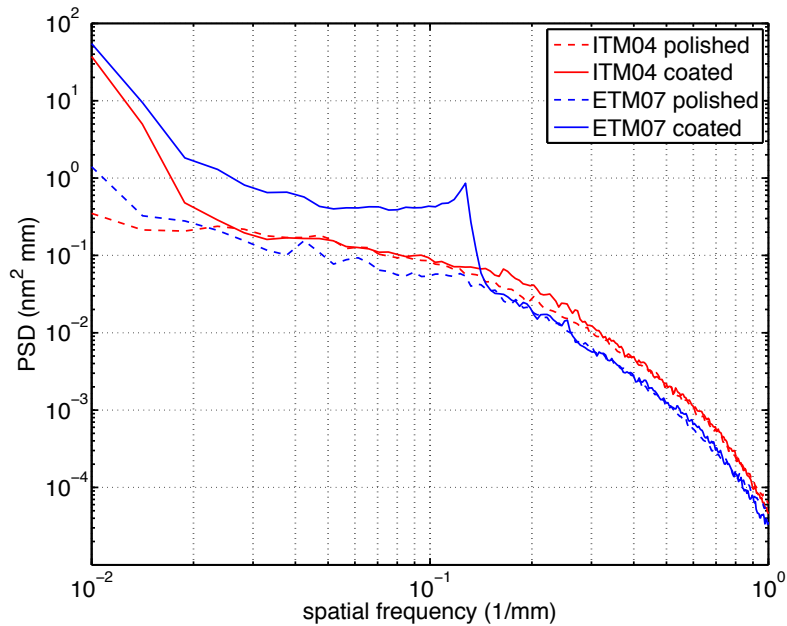

**Figure 7** Power spectra of the surface maps, coated and un-coated, for the ITM and ETM used in one arm of the L1 interferometer.

Though coating absorption is a negligible component of the total loss, it is critical for higher power operation (see section 4.6). Absorption in the test mass high-reflectivity coatings is specified at less than 0.5 ppm, and actual coatings show measured absorption of 0.2—0.4 ppm. Other coatings are specified to have absorption less than 1 ppm.

The test mass coatings delivered to date meet many of their challenging requirements such as absorption and scattering, but some of the properties are not ideal:

- For the ETMs, a systematic error in the layer thicknesses left the 1064 nm transmission in spec, but caused the 532 nm transmission to be larger than desired by an order of magnitude. This has made the arm length stabilization system described in section 4.8.4 much more challenging.
- Features of the coating planetary system produce a ~1 nm peak-to-peak thickness ripple, with an 8 mm period, in the outer regions of the mirror. This is the peak seen in the ETM07 coated curve in Figure 7. This surface ripple scatters light out of the arm cavity; the optical loss is not significant (6 ppm), but the scattered light impinges on the beam tube baffles and can cause excess phase noise (see section 4.7).
- Spherical aberration of the ETM coatings is 2-3 times higher than the specification of 0.5 nm. This will alter slightly the shape of the cavity fundamental mode compared to an ideal Gaussian $TEM_{00}$ mode. However, the aberration is very similar from ETM to ETM, so the arm cavity modes are well matched and this defect does produce significant loss.
- The anti-reflection coatings on the ITMs have 2-3 times higher reflectivity than the specification of 50 ppm. Though not significant in terms of loss, the higher AR produces higher power ghost beams that may need better baffling to control scattered light noise (see section 4.7).

*4.4 Suspensions*

All of the primary in-vacuum interferometer optical components (all depicted in Figure 1, with the exception of the input Faraday isolator) are suspended by pendulum systems of varying designs. These suspension systems provide passive isolation from motion of the seismically isolated optics tables in all degrees of freedom and acceptable thermal noise. The suspensions also provide low noise actuation capability, used to align and position the optics based on interferometer sensing and control signals.

*4.4.1 Requirements.* The suspension designs employed for each optical element depend upon the performance requirements and physical constraints, as listed in Table 6. Most of the suspensions employ multiple pendulum and vertical isolation stages.

**Table 6** Suspension types. Auxiliary suspensions are for optical pick-off beams, beam steering and mode matching.

| Optical Component | vertical isolation stages | pendulum stages | Final stage fiber type | Longitudinal noise requirement @ 10 Hz (m/√Hz) |
|---|---|---|---|---|
| Test Masses (ITM, ETM) | 3 | 4 | Fused silica | $1 \times 10^{-19}$ |
| Beamsplitter (BS) | 2 | 3 | Steel wire | $6 \times 10^{-18}$ |
| Recycling cavity optics | 2 | 3 | Steel wire | $1 \times 10^{-17}$ |
| Input Mode Cleaner (IMC) optics | 2 | 3 | Steel wire | $3 \times 10^{-15}$ |
| Output Mode Cleaner (OMC) Assy | 2 | 2 | Steel wire | $1 \times 10^{-13}$ |
| ETM Transmission Monitor | 2 | 2 | Steel wire | $2 \times 10^{-12}$ |
| Auxiliary suspensions | 1 | 1 | Steel wire | $2 \times 10^{-11}$ |

*4.4.2 Design description.* The most challenging design is the quadruple pendulum suspension for the test masses [27]. Each of these suspensions is comprised of two adjacent chains, each chain having four masses suspended from one another. The main chain includes the test mass optic as the lowest mass. The adjacent, reaction chain provides an isolated set of masses for force reaction. The bottom mass in the reaction chain is the Compensation Plate (CP) optic in the case of an ITM suspension, and the End Reaction Mass (ERM) in the case of an ETM suspension. A structure surrounds and cages the suspended masses and mounts to the seismically isolated optics table.

Vibration isolation for the test mass is accomplished with a 4-stage pendulum and 3 stages of cantilevered blade springs, providing isolation in all 6 degrees-of-freedom above approximately 1 Hz. The suspension is designed to couple 22 of the 24 quasi-rigid body modes (all but the 2 highest frequency) of each isolation chain so that they are observable and controllable at the top mass (4 wires between masses to couple pitch and roll modes; non-vertical wires to couple pendulum modes). The blade springs are made of maraging steel to minimize noise resulting from discrete dislocation movements associated with creep under load [31].

For each chain, all the quadruple suspension rigid body modes below 9 Hz can be actively damped from the top stage. Sensing for this local damping is accomplished with integral optical shadow sensors [32], or with independent optical lever sensors. The shadow sensors are collocated with the suspension actuators and have a noise level of $3 \times 10^{-10}$ m/√Hz at 1 Hz.

Force actuation on the upper three masses is accomplished with coil/magnet actuators [32]. Six degree-of-freedom actuation is provided at the top mass of each chain, by reacting against the suspension structure. These actuators are used for the local damping of 22 modes (each chain). The next two masses can be actuated in the pitch, yaw and piston directions, by applying forces between adjacent suspended masses. These stages are used for global interferometer control. Low noise current drive electronics, combined with the passive filtering of the suspension, limit the effect of actuation noise at the test mass.

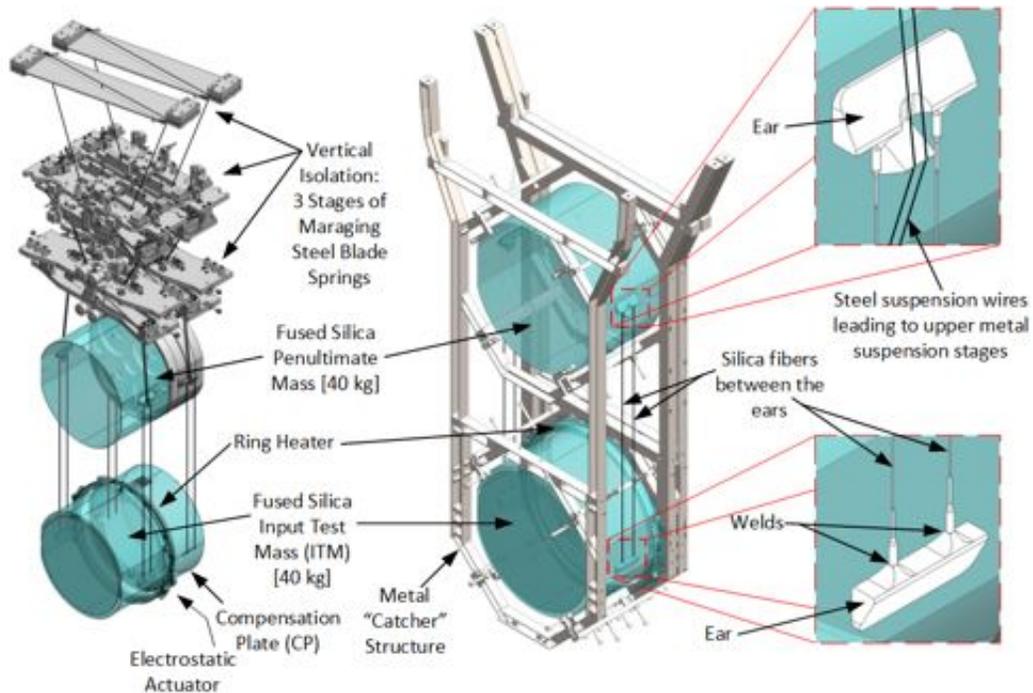

**Figure 8** Quadruple pendulum suspension for the Input Test Mass (ITM) optic.

Direct low-noise, high-bandwidth actuation on the test mass optic is accomplished with electro-static actuation [33]. The CP and ERM each have an annular pattern of gold electrodes, deposited on the face adjacent to the test mass, just outside the central optical aperture. The pattern is separated into 4 quadrants, which enables actuation in pitch, yaw and piston. The force coefficient is highly dependent on the separation between the test mass and its reaction mass. The ETM-ERM separation is 5 mm, which provides a maximum force of about 200 micro-Newtons using a high-voltage driver. Less actuation is needed on the ITMs, so the CP-ITM gap is increased to 20 mm to mitigate the effect of squeezed film damping [23].

The test mass and the penultimate mass are a monolithic fused silica assembly, designed to minimize thermal noise [34]. Machined fused silica elements ("ears") are hydroxide-catalysis (silicate) bonded to flats polished onto the sides of the TM and penultimate mass. Custom drawn fused silica fibres are annealed and welded to the fused silica ears with a $CO_2$ laser system [35]. The shape of the fibres is designed to minimize thermal noise (400 μm dia. by 596 mm long with 800 μm dia. by 20 mm long ends), while achieving a low suspension vertical "bounce" mode (9 Hz, below band) and high first violin mode frequency (510 Hz, above the instrument's most sensitive frequency range). The fibre stress (800 MPa) is well below the immediate, static, breaking strength (5 GPa).

The other suspension types employ the same basic key principles as the test mass quadruple suspensions, except for using steel wire in the final stage.

*4.4.3 Performance.* The transfer functions (actuation to response) in all degrees of freedom in general match very well the model for the test mass suspension; Figure 9 shows one such comparison. Furthermore, the transfer functions are well matched from suspension to suspension, so that very small, if any, parameter value changes are required for the control filters of the multi-input, multi-output control system.

The thermal noise performance of the suspension cannot be verified until the interferometer is fully commissioned and operating at design sensitivity. The displacement thermal noise spectrum for the monolithic stage is made up of quadrature sum of contributions from the horizontal pendulum mode, the vertical mode (with a vertical-horizontal cross-coupling assumed of 0.1%), the first violin mode and the loss associated with the silicate bonded ears. The as-installed quality factor (Q) of the first violin modes of the fibres has been measured to be $1.1$—$1.6 \times 10^9$, which is about a factor of 2 higher than anticipated from prototype testing [34], and is consistent with the anticipated level of thermal noise.

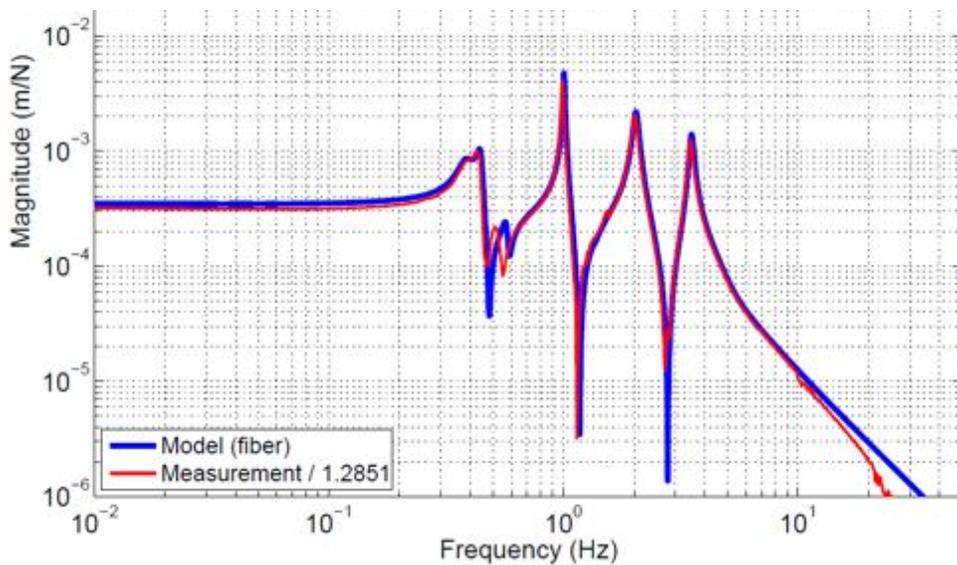

**Figure 9** Transfer function (magnitude) from longitudinal force to longitudinal position of the top mass of an ITM suspension.

*4.5 Seismic isolation*
The general arrangement of the seismic isolation system and its relationship to the vacuum system and the suspension systems is illustrated in Figure 10, where a cut-away view of one of the test mass vacuum chambers is also shown.

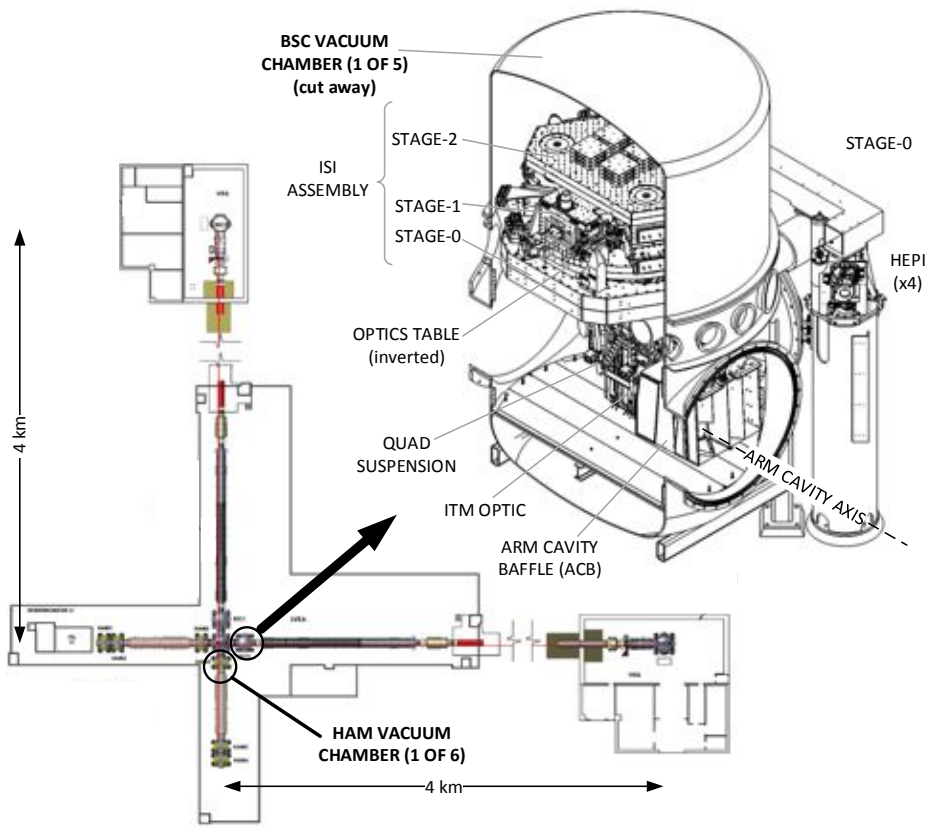

**Figure 10** An Input Test Mass (ITM) vacuum chamber.

*4.5.1 Requirements.* The motion of the detector components, especially the interferometer optics, must be limited to very small amplitudes. The conceptual approach for seismic isolation is to provide multiple stages of isolation, as depicted in Figure 11. The stages topologically closest to the ground are referred to as the seismic isolation system; they provide coarse alignment, and employ both active and passive isolation to deal with the highest amplitude of motion. An optics table provides the interface between the seismic isolation system and the subsequent suspension system. The optics tables in the smaller (HAM) vacuum chambers are 1.9 m by 1.7 m; the downward-facing optics tables in the larger (BSC) vacuum chambers are 1.9 m diameter. The limits to optics table motion (Table 7) are derived from the allowed motion for the interferometer optics using the passive isolation performance of the suspension systems.

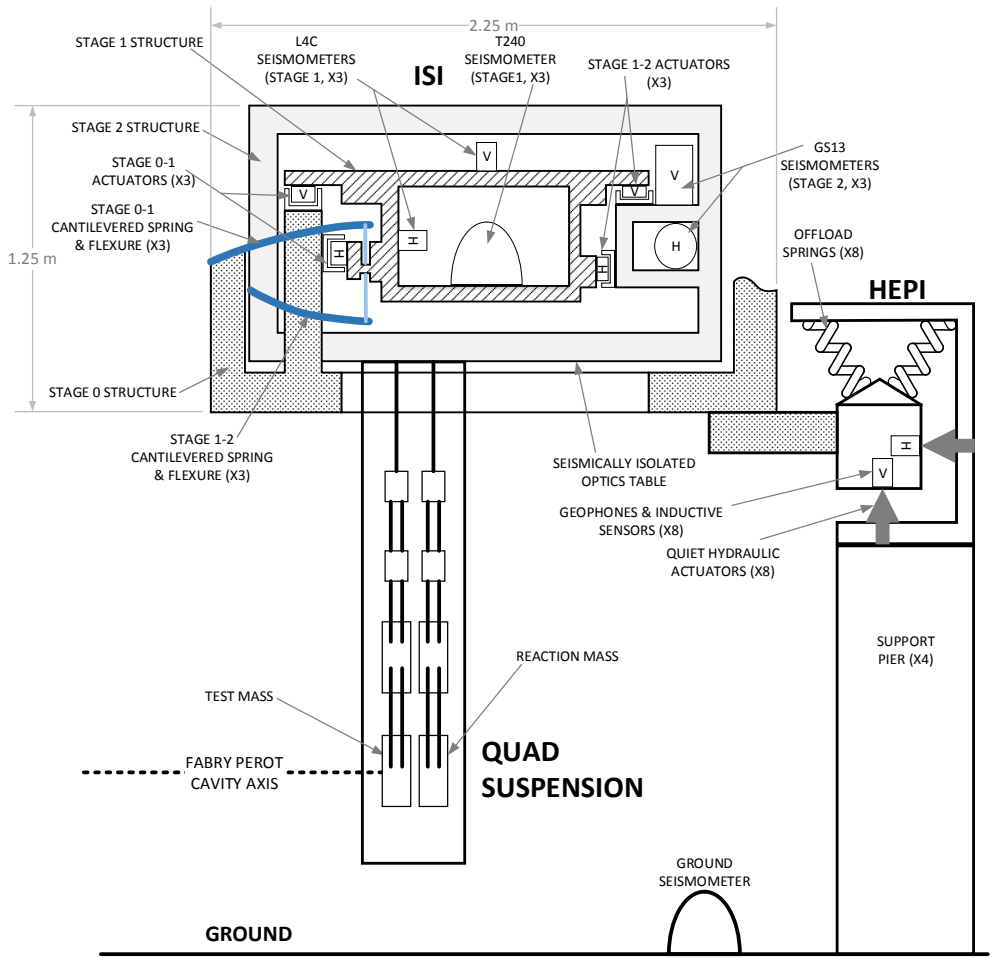

**Figure 11** Seismic isolation for the test mass optic.

**Table 7** Seismic isolation performance requirements for the BSC chamber, 3 stage systems.

| Requirement | value |
|---|---|
| Payload Mass | 800 kg |
| Positioning/alignment range | ± 1 mm |
|  | ± 0.25 mrad |
| Tidal & microseismic actuation range | ± 100 μm |
| Isolation (3 translations) | $2 \times 10^{-7}$ m/√Hz @ 0.2 Hz |
|  | $1 \times 10^{-11}$ m/√Hz @ 1 Hz |
|  | $2 \times 10^{-13}$ m/√Hz @ 10 Hz |
|  | $3 \times 10^{-14}$ m/√Hz @ > 30 Hz |
| Isolation (3 rotations) | $< 10^{-8}$ rad rms, for $1 < f < 30$ Hz |

*4.5.2 Design description.* The first measure taken to reduce ground motion was to site the observatories in seismically quiet locations [36]. The observatory facilities were then designed to limit vibration source amplitudes (e.g., the chiller plant for the HVAC system is remote from the experimental hall). Additional isolation from the already seismically quiet ground motion is provided by the seismic isolation and suspension subsystems.

All in-vacuum interferometer elements are mounted on seismically isolated optics tables (with the exception of some stray light control elements). The typical seismic isolation system consists of two or three stages of isolation (Figure 11). The first stage is accomplished with the Hydraulic External Pre-Isolator (HEPI) system, external to the vacuum system. The next one or two stages are referred to as the Internal Seismic Isolation (ISI) system and are contained within the vacuum system. The test mass, beamsplitter and transmission monitor suspensions are supported by inverted optics tables which have two in-vacuum stages, housed in the BSC chambers. All other interferometer elements are supported by upright optics tables connected to one-stage ISI systems, housed in smaller (HAM) vacuum chambers.

The in-vacuum payload is supported by a structure that penetrates the vacuum chamber at four locations, through welded bellows. Each of these four points is supported by a HEPI assembly; each HEPI system supports a total isolated mass of 6400 kg. HEPI employs custom designed, laminar flow, quiet hydraulic actuators (8 per vacuum chamber) in a low frequency (0.1—10 Hz), 6 degree-of-freedom active isolation and alignment system. The actuators employ servo-valves in a hydraulic Wheatstone bridge configuration to control deflection of a diaphragm by differential pressure. For sensors, HEPI uses a blend of geophones and inductive position sensors. In addition, a ground seismometer provides a signal for feed-forward correction. The HEPI system was deployed for the initial LIGO L1 interferometer [37] and remains essentially the same for Advanced LIGO.

The ISI [38] consists of three stages (each a stiff mechanical structure) that are suspended and sprung in sequence: stage 0 is the support structure connected to the HEPI frame; stage 1 is suspended and sprung from stage 0; stage 2 is suspended and sprung from stage 1. The elastic, structural modes of stages 1 and 2 are designed to be > 150 Hz in order to keep them high above the upper unity gain frequency (~40 Hz) of the control system. The stage 2 structure includes the optics table upon which payload elements, such as the suspensions, are attached. Each suspended stage is supported at 3 points by a rod with flexural pivot ends, which are in turn supported by cantilevered blade springs. The springs provide vertical isolation and the flexure rods provide horizontal isolation. They are both made of high strength maraging steel in order to reduce noise resulting from discrete dislocation movement events associated with creep under load. The springs and flexures allow control in all 6 rigid body degrees-of-freedom of each stage, through the use of electromagnetic force actuation between the stages. Passive isolation from base (stage 0) motion is achieved at frequencies above the rigid body frequencies of these stages.

Stage 1 of the ISI is instrumented with 6 capacitive position sensors (MicroSense), 3 three-axis seismometers (Nanometric Trillium T240) and 6 geophones (Sercel L4C). Stage 2 is instrumented with 6 capacitive position sensors (MicroSense) and 6 geophones (Geotech GS13). The single-axis sensors are equally divided into horizontal and vertical directions. All of the inertial sensors are sealed in vacuum-tight canisters. The 12 electromagnetic actuators (6 between stages 0-1 and 6 between stages 1-2) are equally split between vertical and horizontal orientations, and are a custom, ultra-high vacuum compatible design.

The capacitive position sensor signals provide positioning capability, and are low-pass filtered in the isolation band. For stage 1 control, the T240 and L4C seismometer signals are blended together to provide very low noise and broadband inertial sensing; they are high-pass filtered to remove sensor noise and other spurious low frequency signals, such as tilt for the horizontal sensors. The T240 signals are also used for feed-forward to the stage 2 controller. Before filtering, the sensors are transformed into a Cartesian basis by matrix multiplication in the multi-input, multi-output digital control system.

*4.5.3 Performance.* An example of the isolation performance achieved to date for the three-stage seismic isolation system is shown in Figure 12.

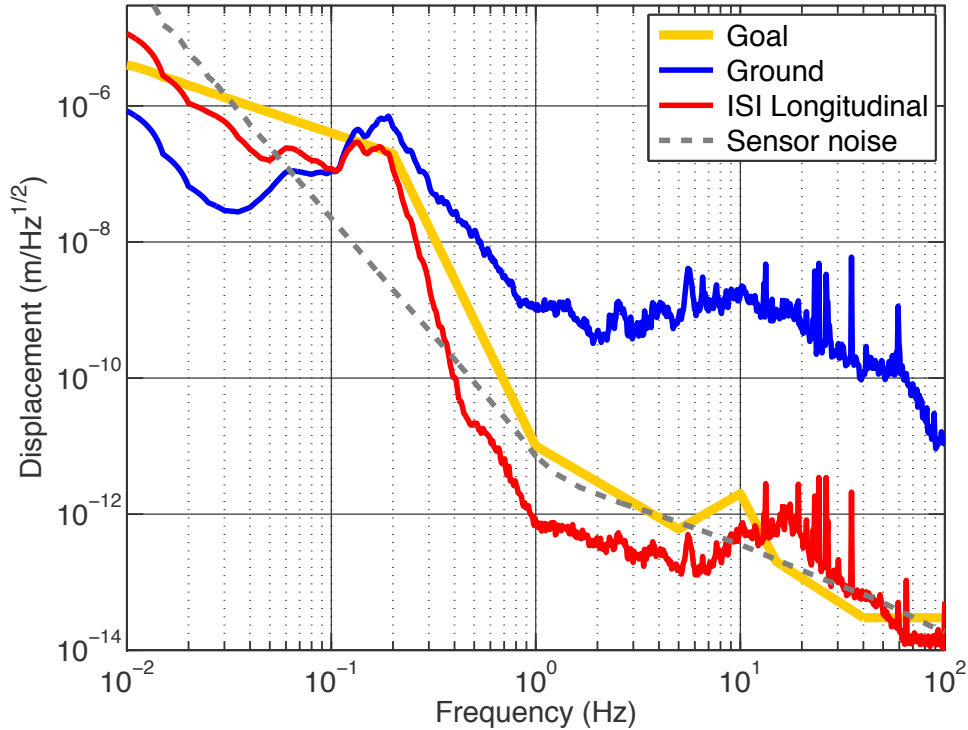

**Figure 12** Example of the BSC ISI optics table isolation performance (from LHO). The blue curve shows the horizontal ground motion adjacent to the vacuum chamber. The gold curve shows the goal performance. The red curve is an in-loop measurement of the optics table longitudinal motion, using the stage 2 inertial sensors. At the mid-frequencies, these inertial sensor signals are pulled below their intrinsic noise floor by the feedback loop. Below ~30 mHz, the horizontal inertial sensors are likely contaminated by tilt coupling.

*4.6 Thermal compensation*

Absorption of the Gaussian-profiled laser beam in the core optics causes a non-uniform temperature increase leading to wavefront aberration through the thermo-optic and thermo-elastic effects (elasto-optic effects are negligible). While thermal aberrations will occur in all the optics to some extent, only the test mass optics are anticipated to require active compensation.

The projected thermal effects, without compensation, are indicated in Table 8. The only place of significant power absorption is at the test mass high-reflectivity surfaces, and the largest optical distortion is the resulting thermal lensing in the test mass substrate. The ITM thermal lensing affects primarily the RF sideband mode in the recycling cavities; the carrier mode is enforced by the arm cavities and is much less perturbed by the ITM lensing. The test mass surface distortion due to thermo-elastic expansion is about 10x smaller than the substrate lensing distortion, but at high power it becomes significant. Uncompensated, the surface distortion would increase the ITM and ETM radii of curvature by a few tens of meters, which would reduce the beam size on each test mass by about 10%. This would increase coating thermal noise, and reduce the coupling efficiency between the arm cavity and recycling cavity modes.

**Table 8** Absorbed power and corresponding thermal distortion at full power: 750 kW in each arm cavity, 5.2 kW in the PRC. Assumes absorption of 0.5 ppm at each HR surface. The distortion values are peak-to-valley over the beam diameter ($1/e^2$). The undistorted sagitta of a test mass is 725/850 nm (ITM/ETM).

| Element | Absorbed power | Surface distortion | Lensing optical path distortion |
|---|---|---|---|
| Test mass HR surface | 375 mW | 15 nm | 190 nm |
| ITM & CP substrates | < 32 mW | < 0.4 nm | < 20 nm |
| BS 50/50 surface | 2.6 mW | 0.1 nm | 1.3 nm |
| BS substrate | < 8 mW | < 0.1 nm | < 4 nm |

*4.6.1 Requirements.* The TCS is required to:
- adjust the arm cavity spot size by adding up to 30 micro-diopters of power to all test mass HR surfaces,
- compensate thermal aberrations in the recycling cavities sufficiently to maintain adequate RF sideband power buildup,
- maintain recycling cavity and arm cavity mode overlap to within 95%, and
- maintain the contrast defect ratio by limiting the additional DC power at the output mode cleaner (OMC) to within 1 mW

The TCS capability to alter the test mass radius-of-curvature can also be used to adjust the transverse optical mode spacing and potentially control optical-acoustic parametric instabilities at high optical power.

*4.6.2 Design description.* The TCS consists of three major elements (see Figure 13): a radiative ring heater (RH), a $CO_2$ laser projector (CO2P) and a Hartmann wavefront sensor (HWS). The RH is used to correct surface deformation of the ITM and ETM, and to partially correct the ITM substrate thermal lens. The CO2P is used to compensate residual ITM substrate distortions not corrected by the RH. The HWS is used to measure the test mass thermal aberrations.

The basic intent of the combination of the RH and CO2P is to add heat to form the conjugate aberration to the thermal lens formed by the main beam heating. Whereas the RH power stability is good enough to permit direct actuation on the test mass, the CO2P actuates on the compensation plate to limit the effects of $CO_2$ laser noise.

The RH assembly is comprised of nichrome heater wire wound around two semi-circular glass rod formers, which are housed within a reflective shield; the inner radius of the assembly is 5 mm larger than the test mass radius. The nichrome wire spacing along the glass former is variable in order to partially compensate for boundary condition effects and achieve better uniformity. The RH is mounted around and radiates onto the test mass barrel. It is positioned near the test mass anti-reflection face so that the thermal flexure of the optic produces a concavity of the high-reflectivity face, correcting the convex deformation caused by the main beam heating. The thermal flexure is very closely approximated by a spherical curvature at the high-reflectivity face. In addition, the thermal lensing produced by the RH partially compensates the lensing produced by the main beam. The effectiveness of the corrections can be expressed by considering a $TEM_{00}$ mode probe beam that passes through or reflects from the optics; the correction factor is defined as the power scattered out of the $TEM_{00}$ mode by the distorted optic, relative to the scattered power for the compensated optic. The RH provides a correction factor of about 10 for both types of test mass distortion (surface and substrate).

The CO2P is configured to make static pattern corrections via both central and annular projected heating patterns. These two patterns are created using masks that are inserted into the beam using flipper mirrors. The annular pattern augments and refines the RH correction, providing another factor of 10 or more correction of the substrate lensing. The central pattern can be used to maintain central heating upon lock loss, for faster return to operation; it can also be used to correct non-thermal power terms due to substrate inhomogeneity. The power is adjusted with a polarizer and rotatable ½-waveplate; power fluctuations are stabilized using an AOM. The CO2P is designed to deliver at least 15W to the compensation plate.

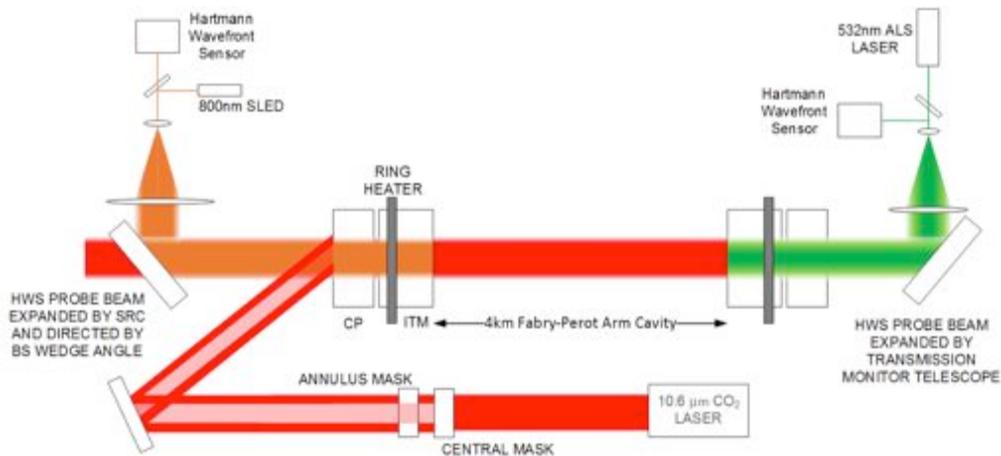

**Figure 13** Schematic layout of the Thermal Compensation System for the X-arm. A similar configuration is implemented on the Y-arm, with the SLED probe beam instead transmitting through the BS. Each of the masks can be independently flipped in or out of the beam path.

The HWS is used to measure the thermal distortions, both before and after compensation. The HWS can measure thermal wavefront distortions with a sensitivity of < 1.4 nm and a spatial resolution of ≤1 cm, both over a 200 mm diameter at the ITM. In the vertex, the HWS uses super-luminescent LEDs (SLED) for probe beams. Independent probe beam paths for the X and Y-arms of the Michelson are achieved through choice of SLED wavelength, given the spectral properties of the beam splitter coating: 800 nm is used for the X-arm and 833 nm for the Y-arm (5 mW for both). Cross-talk between the X and Y paths is less than 1 part in 100. Both probe beams make use of the beam expansion telescope integral to the signal recycling cavity (SR2 & SR3) to achieve a large profile at the BS, CP and ITM. The HWS for the ETM uses the ALS green laser beam as a probe.

*4.7 Scattered light control*

A significant fraction of the input laser light ends up outside the interferometer mode through scattering or reflections from the various optics. This light will hit and scatter from surfaces that are typically not as well mechanically isolated as the suspended optics, picking up large phase fluctuations relative to the main interferometer light. Thus even very small levels of scattered light can be a noise source if it rejoins the interferometer mode. Scattered light phase noise is proportional to the electric field amplitude, and for small surface motion proportional to the amplitude of motion. For large motion amplitude, fringe wrapping occurs, resulting in upconversion [39].

The severity of a potential light scattering surface depends upon the location within the interferometer. Light scattered into a single arm cavity, or into the anti-symmetric port, directly contributes to an apparent differential signal. Light scattered into the power recycling cavity, on the other hand, is less impactful due to the common mode rejection of the arm cavities.

The basic design approach for stray light control [40] is to (a) capture all first-order ghost beams in beam dumps, (b) baffle all views of the vacuum envelope (which is not isolated from ground motion), (c) provide baffles and apertures with low light scattering surface properties (low bidirectional reflectance distribution function (BRDF) surfaces), (d) provide baffle geometries which trap specular reflections into multiple reflections, (e) provide surfaces with some amount of absorption, and (f) vibration isolate the baffles. Surface treatments are restricted due to the need for ultra-high vacuum compatibility. The larger baffles and apertures are oxidized, polished stainless steel surfaces. The depth of the oxidized layer must be carefully controlled to maintain a well-bonded oxide layer and prevent a frangible layer. Smaller baffles and beam dumps are made of black glass. High power, in-vacuum, beam dumps are comprised of polished, chemical vapour deposited silicon carbide.

The locations of the primary baffles are shown in **Error! Reference source not found.**. An example of one of the baffle configurations is the Arm Cavity Baffle (ACB). This baffle has a "W" cross-section with an aperture slightly larger than the test mass diameter. A version of this baffle is placed in the arm cavity in front of each test mass optic. The small angle scatter from the far test mass (4 km away) is caught on the beam tube side of the ACB, and forced into multiple reflections in the acute angle 'cavities' formed by the baffle surfaces. Large angle scatter from the adjacent test mass is caught on the other side of this suspended baffle. The ACB is mounted with a single pendulum suspension (1.6 Hz), with a cantilevered spring for vertical isolation (2 Hz).

Light scattered off the test masses at even smaller angles is caught by the approximately two hundred baffles mounted along the beamtube. These baffles were installed for initial LIGO, but were designed to be effective for later generations of interferometers.

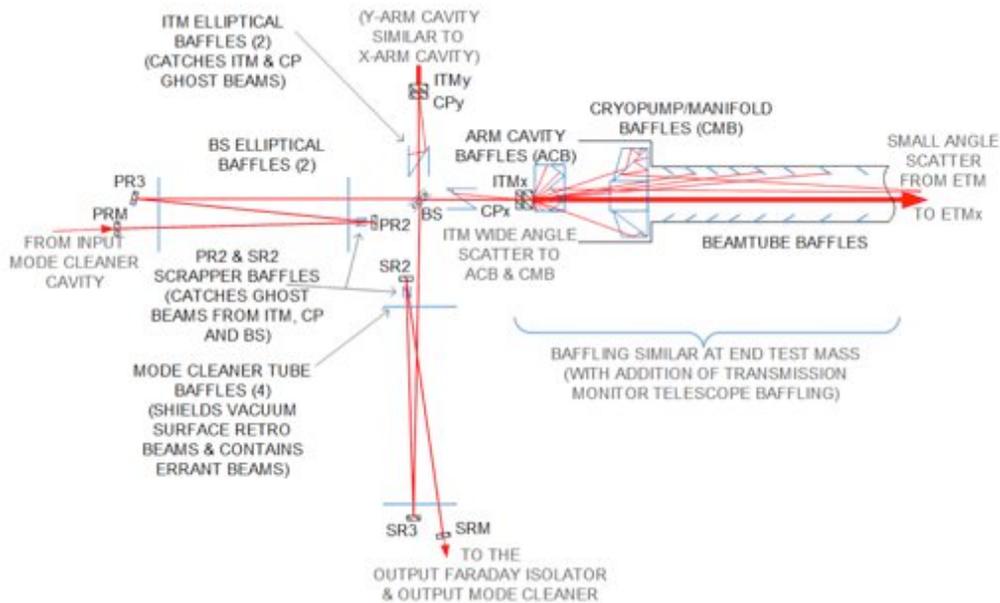

**Figure 14** Locations of the primary scattered light baffles.

*4.8 Global sensing and control*

Using primarily interferometrically generated error signals, active feedback is used to keep the interferometer at the proper operating point. This entails keeping the four interferometer cavities on resonance (two arm cavities, power- and signal-recycling cavity), and the Michelson at the dark fringe (or at a controlled offset from there). In addition, global controls are required to keep the whole interferometer at the proper angular alignment.

The optical sensing for both the length and alignment degrees-of-freedom is derived from photodetectors located at five output ports of the interferometer:

- Reflection port (**REFL**): the light reflected from the power recycling mirror; typically ~10% of the beam is detected
- Power recycling cavity pick-off (**POP**): the beam transmitted by PR2 coming from the beamsplitter; typically ~10% of the beam is detected
- Anti-symmetric port (**AS**): the beam exiting the signal recycling mirror; typically 99% of this beam is directed to the DC readout detectors, and 1% is directed to alignment and auxiliary detectors
- ETM transmissions (**TRX & TRY**): beams transmitted by the two end test masses; typically a few percent of each beam is detected

*4.8.1 Length sensing and control.* The length sensing scheme is similar to that used in initial LIGO [41], extended to sense the length of the new signal recycling cavity. As per standard practice, the arm cavity lengths are treated in the basis of common and differential modes: the former is the average arm length, equivalent to the laser frequency, and the latter is the arm length difference, which is also the gravitational wave signal mode.

Two sets of RF modulation sidebands are applied to the input laser field: one at 9 MHz and one at 45 MHz. Both pairs of RF sidebands are resonant in the power-recycling cavity, but not in the arm cavities. The Michelson contains the usual Schnupp asymmetry so that RF sideband power is transmitted to the AS port even when the carrier is at the dark fringe. In this case, we choose the asymmetry to couple the majority of the 45 MHz sideband power into the signal recycling cavity. This can be achieved either with a relatively small (several cm) or relatively large (tens of cm) asymmetry, but the former is preferred because it suppresses the 9 MHz power in the SRC, leading to better separation of error signals for the two recycling cavities. With our Schnupp asymmetry of 8 cm, and a SRM transmission of 35%, nearly all of the 45 MHz sideband power is transmitted to the AS output, while only about 0.3% of the 9 MHz power arrives there.

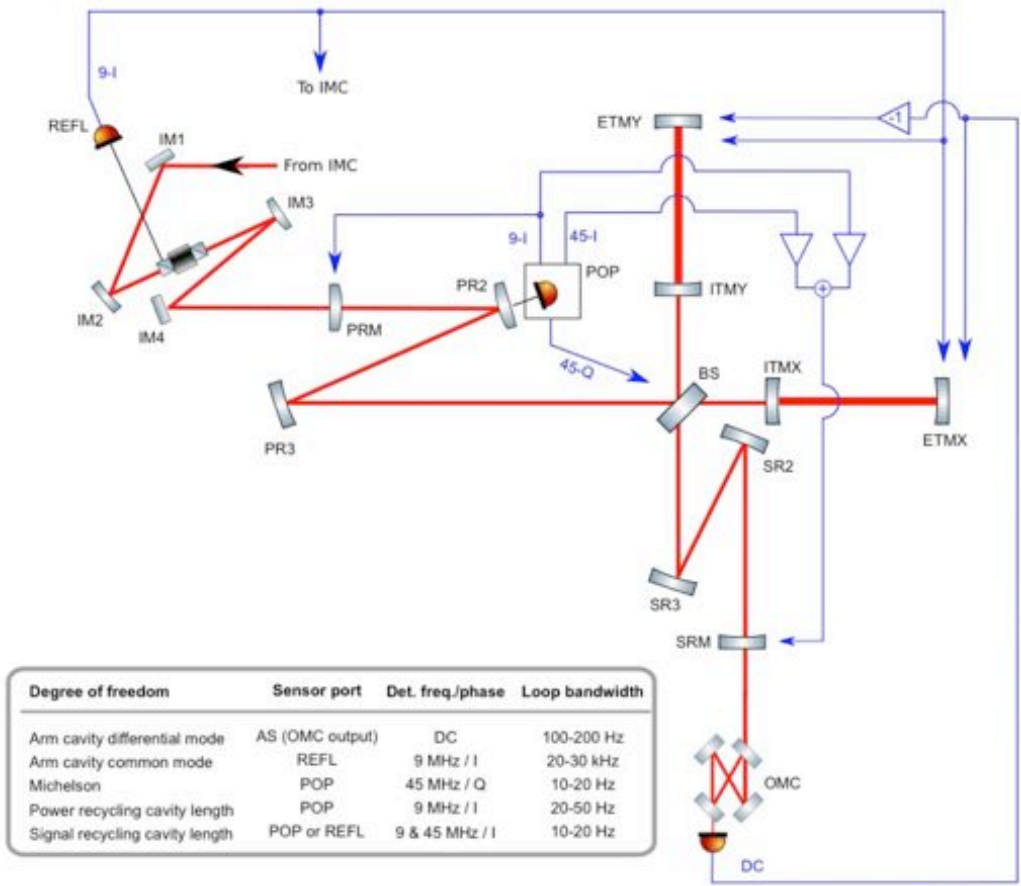

**Figure 15** Length sensing and control scheme for Advanced LIGO. Though not shown, the Michelson signal is also fed back to the PRM and SRM, so that the BS actuation affects only the Michelson degree-of-freedom.

Figure 15 summarizes the sensing and control scheme for the length degrees-of-freedom. Except for the DC readout of the GW channel, the length signals are derived from photodetectors at the REFL and POP ports, by demodulating their outputs at one or more of the RF modulation frequencies. The feedback controls are implemented digitally, with a real-time sampling rate of 16,384 samples/sec. The arm common mode loop actuates on the laser frequency so that it follows the highly stable common arm length. The servo includes an analogue feedback path to achieve high bandwidth. The design of this complex common mode servo follows that of initial LIGO [1].

All of the photodetectors used for the length sensing during low-noise operation are located in vacuum chambers, mounted on seismically isolated platforms. This is done to reduce spurious noise effects in the signal detection, such as from vibrations of the detectors, acoustic modulation of paths in air, or from particulates falling through the beams.

For the auxiliary degrees-of-freedom (DoF) – all those other than the arm cavity differential mode – the control loop design involves a trade-off between having adequate loop gain to avoid non-linear effects, and limiting the coupling of noise into the GW channel. These auxiliary DoF couple weakly to the GW channel, at a level of order $10^{-2}$ or smaller (meters/meter). However, the shot-noise limited sensing of these channels is 3-5 orders of magnitude worse—in terms of equivalent displacement—than the GW channel. Thus if the sensing noise of the auxiliary channels is impressed onto these DoF through their feedback loops, it can easily spoil the GW sensitivity. This noise mechanism is mitigated partially through the design of the auxiliary DoF servo loops: the loop bandwidths are kept low, and low-pass filters with cut-offs just above the unity gain frequencies are employed. In addition, *noise correction* paths are implemented: the auxiliary loop control signals are appropriately scaled and filtered, then added to the GW channel control signal to remove their coupling (feed-forward noise suppression). These correction paths reduce the auxiliary DoF noise infiltration by 30-40 dB.

*4.8.2 Alignment sensing and control.* The residual angular motion of the arm cavity mirrors must be 1 nrad-rms or less to adequately suppress alignment noise effects [42]. This is 2-3 orders of magnitude smaller than their locally-controlled angular motion. Somewhat less stringent angular stability requirements apply to the other interferometer optics. The alignment sensing and control scheme designed to provide this level of stability is an extension of the Initial LIGO design [43]. The same phase modulation-demodulation techniques used for the length sensing are used with quadrant photodiodes to produce an alignment wavefront sensor (WFS). A WFS produces an alignment signal via the interference between the fundamental $TEM_{00}$ Hermite-Gaussian field component and the $TEM_{01}/TEM_{10}$ field components generated by misalignments. Several new aspects of the alignment control design must be considered for Advanced LIGO:

- High circulating power in the arms produces significant radiation pressure torques on the cavity mirrors
- Additional degrees-of-freedom: signal recycling introduces a new mirror that must be aligned; the folded recycling cavities introduce another four mirrors
- Alignment control noise must be filtered out starting at a much lower frequency

The radiation pressure effects are typically described in terms of a *soft* and *hard* alignment mode, depending on whether the optical torque subtracts from or adds to the mechanical stiffness of the suspension [44]. This is illustrated in Figure 16 for a single cavity. In Advanced LIGO, the radiation pressure torque becomes greater than the suspension mechanical torque at arm cavity powers above 700 kW. Above this, the soft mode becomes dynamically unstable, and must be actively stabilized. As the unstable eigenfrequency stays low even at full power (less than 0.3 Hz), it is straightforward to stabilize even with low bandwidth.

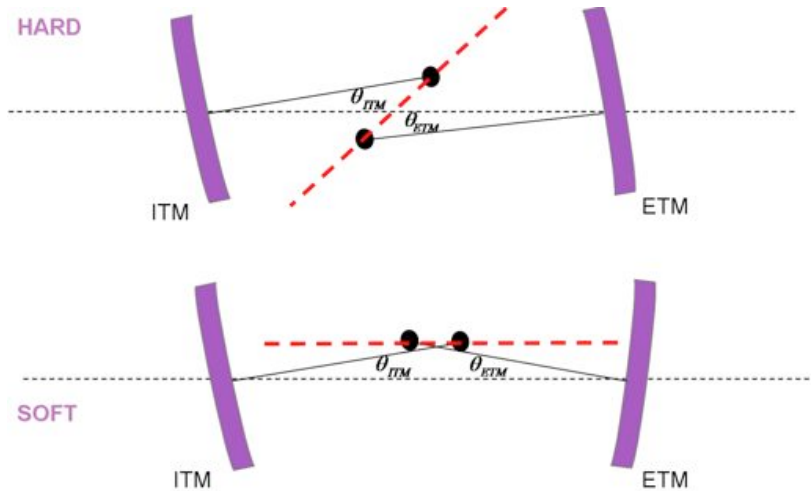

**Figure 16** Representation of the hard and soft modes of cavity mirror alignment.

The alignment WFS are located at the REFL and AS interferometer ports; two WFS are installed at different beam line positions at each port, such that they are separated in Gouy phase by 90 degrees. This allows combinations of WFS signals to be formed to reconstruct any Gouy phase. In addition to the WFS, several quadrant photodiodes (QPDs) provide relative beam position information. The TRX, TRY and POP ports each have two QPDs, with each pair separated by 90 degrees Gouy phase. Lastly, the AS port contains a single QPD (detecting a small fraction of the AS beam). As with the length sensors, all of the alignment sensors that are used in low-noise operation (both WFS and QPD) are located in vacuum chambers, mounted on seismically isolated platforms.

The control scheme for the arm cavity mirrors and the beamsplitter is summarized in Table 9. The other optics can be controlled using various combinations of the other WFS and QPD signals; these loops are implemented essentially for drift control, with bandwidths of 0.1 Hz or less.

**Table 9.** Alignment control scheme for the arm cavity mirror (test masses) and the beamsplitter.

| Degree of freedom | Sensor type & port | Detection frequency/phase | Loop unity gain frequency |
|---|---|---|---|
| Arm cavity differential hard mode | WFS AS | 45 MHz Q-phase | 1-few Hz |
| Arm cavity common hard mode | WFS REFL | 9 MHz I-phase | ~1 Hz |
| Arm cavity differential soft mode | QPD TRX-TRY | DC | ~1 Hz |
| Arm cavity common soft mode | QPD TRX+TRY | DC | ~1 Hz |
| Beamsplitter | WFS AS | 36 MHz Q-phase | ~0.1 Hz |

*4.8.3 Output mode cleaner and DC readout.* The output mode cleaner (OMC) is designed to filter out all RF sideband and higher-order spatial mode light at the AS port, so that the main photodetectors receive only light that carries the gravitational wave signal. The OMC is a bowtie cavity with a moderate finesse (400); this is a trade-off between having a narrow bandwidth to filter out RF components, and maintaining a high transmission efficiency (> 95%). The OMC higher-order mode spacing is fine-tuned (via the cavity length) to avoid overlap with higher-order modes likely to exit the interferometer AS port. The nominal OMC round trip length is 1.13 m.

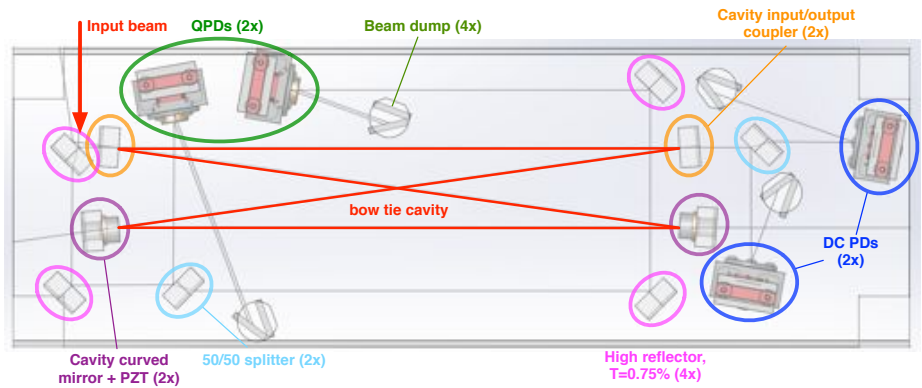

**Figure 17** Layout of the OMC on the fused silica breadboard (dimensions 45cm x 15cm x 4.1cm).

As shown in Figure 17, the OMC cavity optics and the output photodiodes are all bonded to a single breadboard of fused silica. This minimizes relative motion between the OMC output beam and the diodes, which can otherwise be a source of noise. The breadboard also includes two QPDs for aid in alignment, and multiple beam dumps for scattered light. Two of the OMC mirrors are mounted on PZT actuators for length control of the cavity. The OMC cavity is locked to the AS port beam with a dither scheme: one of the PZTs applies a small modulation to the cavity length at a frequency of several kHz; the output detectors are demodulated at the dither frequency, and the resulting error signal is fed back to the other PZT to maintain resonance. For vibration isolation, the whole breadboard is mounted in a two-stage suspension that has active damping and pointing capability. The OMC suspension is mounted in a HAM vacuum chamber on a HAM seismic isolation platform.

The AS port beam is directed into the OMC by three steering optics. These optics provide mode-matching to the OMC, and they are mounted in active single-stage suspensions for vibration isolation and pointing control. There are several possible schemes for alignment optimization and control of the beam into the OMC [45]. A simple method is to dither the pointing direction of the AS port beam using the suspended steering optics, and demodulate the OMC transmitted signal at the dither frequencies.

*4.8.4 Arm length stabilization and lock acquisition.* Lock acquisition is the process of bringing the interferometer to its operating point, with all the cavities resonant so that power buildup is at a maximum. Advanced LIGO implements a new lock acquisition scheme that incorporates two significant features:

- An arm length stabilization system [46] is used to control the microscopic length of each arm cavity, independently of the other degrees-of-freedom
- The vertex dual-recycled Michelson interferometer is initially locked with the third harmonic technique [47] (3f-technique), providing more robust control during the acquisition process

The idea of the arm length stabilization system (ALS) is to lock each arm cavity individually using a laser mounted behind each end test mass. The arm cavities are then held away from resonance for the main laser light by offsetting the frequency of the end station ALS lasers, relative to the main laser frequency. The ALS lasers are doubled Nd:YAG lasers, operating at 532 nm to distinguish them from the main laser light; they also have a 1064 nm output which is used to phase (frequency offset) lock them to the main laser light. The test masses have dichroic coatings designed to create a relatively low finesse cavity for 532 nm, with reasonable transmission of 532 nm light to the vertex beamsplitter.

In the vertex, one of the ALS beams is interfered with a doubled sample of the main laser light, and the two ALS beams are interfered with each other (see Figure 18). The former produces a common mode signal that is used to control the frequency of the main laser. The latter produces a differential mode signal that is used to stabilize the differential arm length.

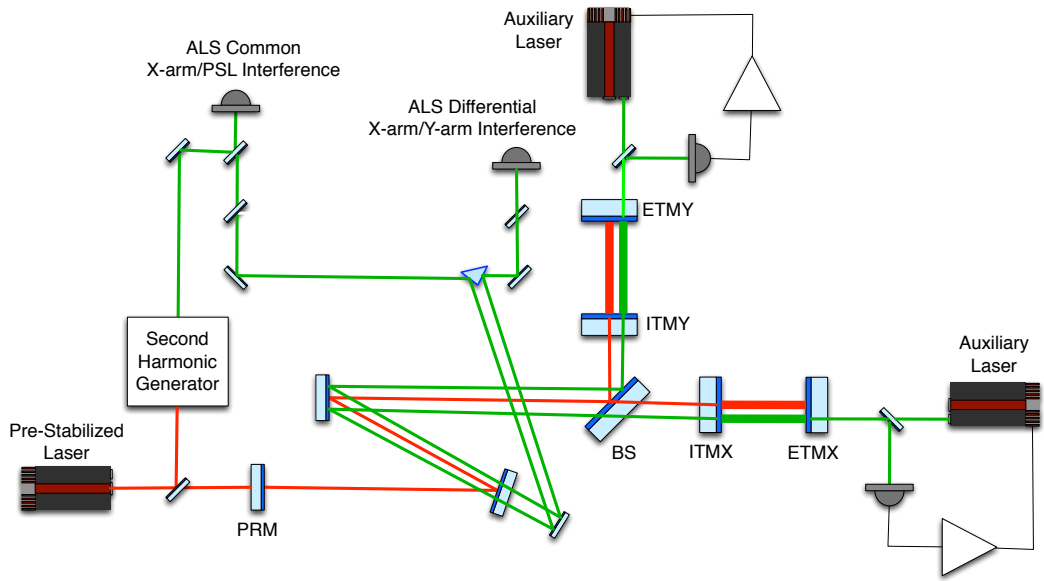

**Figure 18** Simplified schematic of the arm length stabilization system.

The acquisition process is as follows [48]: Each ALS laser is locked to its arm cavity at 532 nm. The ALS common and differential mode loops are then engaged, using the vertex 532 nm heterodyne signals. The common mode frequency difference is set so that the main 1064 nm laser light is about 500 Hz away from resonance in the arm cavities. The vertex dual-recycled Michelson is then locked using the 3f-technique. Next, the common mode frequency offset is gradually reduced to zero to bring the interferometer to the operating point; this step is the most critical and in fact involves several transitions that are fully described in reference [48]. At the operating point, all sensor signals are switched over to the low-noise, 1064 nm signals described in section 4.8.1.

*4.9 Calibration*

The output signal of the LIGO detector is a calibrated time series of "external" differential length variation between two arm cavities, reconstructed from control and error point signals of the differential arm length servo by applying an appropriate response function.

Errors in the interferometer response function degrade the ability to detect gravitational waves and the ability to measure source properties of detected signals [49]. Calibration accuracy is naturally more important for source parameter extraction than for detection. We have set the calibration accuracy requirements at 5% in amplitude and 16μsec in timing, over 2σ confidence levels. This is consistent with requirements for detection of strong binary black-hole signals [49].

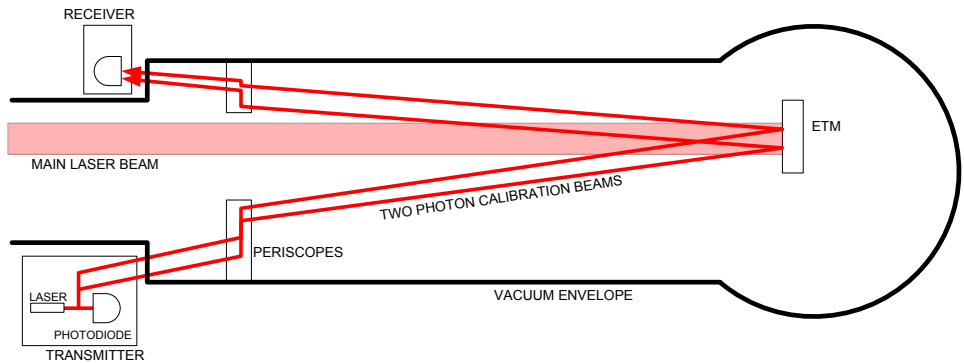

**Figure 19** Schematic of the photon calibration system.

Three different calibration methods are employed to allow cross-checks on accuracy: (a) free-swinging Michelson [50]; (b) frequency modulation [51]; (c) photon calibrator [52]. In the free-swinging Michelson method the interferometer laser light is used as the length reference to calibrate Michelson interference fringes as the test masses either move freely or are forced through several microns of motion. This method calibrates ETM actuation with drive amplitude on the order of $10^{-12}$ m. The frequency modulation method is used with a single resonant arm cavity. In this configuration, a known modulation of the laser frequency calibrates the cavity sensing signal in terms of equivalent length modulation. A simultaneous drive of the ETM actuator ($10^{-14}$ m) at a slightly different frequency is then used to derive its actuation calibration.

The photon calibration method uses an auxiliary, power-modulated laser to create displacements via photon recoil off the surface of the ETM (Figure 19). Two beams are used, offset symmetrically from the mirror center. The offset moves elastic deformation caused by the photon pressure away from the location of the main interferometer beam, and the two-beam geometry avoids imposing torque. The laser power is measured with a temperature stabilized InGaAs photodetector mounted on an integrating sphere, which is calibrated against a NIST-calibrated standard. The lasers are 2 W continuous-wave, 1047nm Nd:YLF lasers. An AOM enables power modulation, up to a peak-to-peak sinusoidal modulation of 1 W, producing an ETM displacement of $10^{-16}$ m-rms at 100 Hz.

*4.10 Physical environmental monitoring*
To complete each LIGO detector, the interferometers described so far are supplemented with a set of sensors to monitor the local environment (Figure 20). Seismometers and accelerometers measure vibrations of the ground and various interferometer components; microphones monitor acoustic noise at critical locations; magnetometers monitor fields that could couple to the test masses or electronics; and radio receivers monitor RF power around the modulation frequencies. This is an expanded version of the environmental monitoring system employed in initial LIGO [53]. These sensors are used to detect environmental disturbances that can couple to the gravitational wave channel.

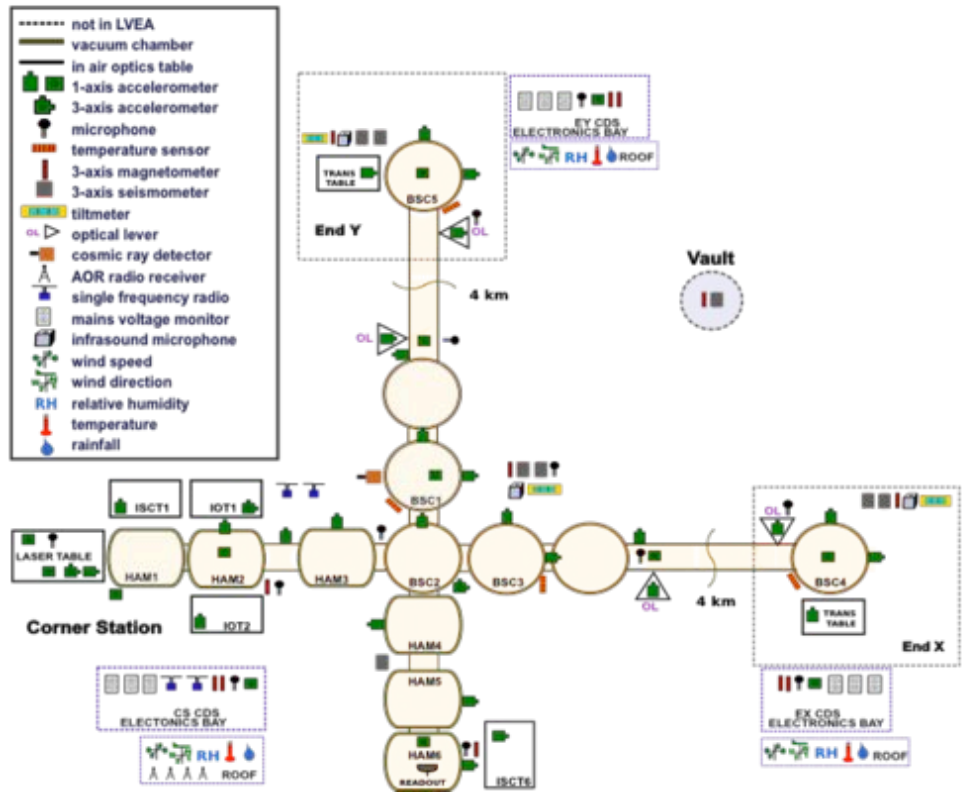

**Figure 20** Physical environmental monitoring sensors at LLO. The test masses are in the vacuum chambers BSC1/3 (ITMs) and BSC4/5 (ETMs), and the 50/50 beamsplitter is in BSC1.

*4.11 Controls and data acquisition*
Many of the subsystems employ real-time digital controls in their functioning; data from other subsystems and channels must be acquired in real-time for archiving and analysis.

A custom data acquisition architecture, diagrammed in Figure 21, is implemented to serve these needs.

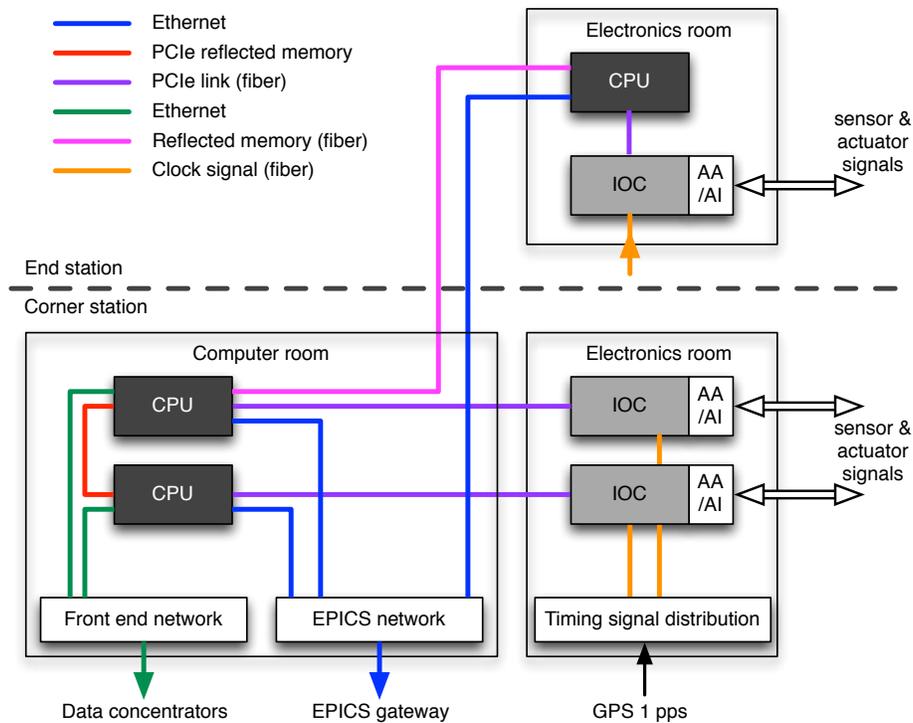

**Figure 21** Schematic of the data acquisition architecture.

Interferometer sensor and actuator signals flow through custom input-output chassis (IOC). Each IOC houses: a 17 slot PCI Express (PCIe) backplane; a timing module that provides accurate triggering at 65536 Hz and is synchronized with the interferometer timing distribution system; a commercial fibre optic PCIe uplink to a real-time control computer. The PCIe slots are populated specifically for each IOC, but primarily contain commercial multi-channel, simultaneous sampling 16-bit analogue-to-digital and 16 and 18-bit digital-to-analogue converters (ADCs and DACs). Custom anti-alias and anti-image filters interface between the ADCs and DACs and the analogue signals. The IOCs and filters are located in rooms separate from the vacuum chamber areas to mitigate electronic and acoustic interference with sensitive interferometer components.

The real-time control computers are located yet further from the interferometer, in a separate computer room, and are linked via fibre. The computers are commercial, multi-core rack mount units, specially selected for compatibility with real-time operation: there must be no uncontrollable system interrupts, and they must be capable of supporting the required number of PCIe modules.

The system is designed to support servo loop rates of up to 65536 Hz, which requires real-time execution to be precise and repeatable to within a few microseconds, and synchronized across the site. To attain this performance, the GPL Linux operating system is used on the control computers, but with a LIGO custom kernel patch. This patch allows the isolation of a given CPU core from the Linux system for exclusive use by a real-time control program.

Specific subsystem data acquisition and control programs are created with the aid of a real-time code generator (RCG). The RCG provides a set of custom software modules, as well as support of many MATLAB Simulink parts. Using Simulink as the GUI, code is developed as a graphical model that describes the desired execution sequence. Standard scripts then produce executable software from the model.

A key RCG component is the IIR (infinite-impulse response) filter module, which is used to define control transfer functions and channel calibrations. A given control model may contain a few hundred of these modules, each module containing up to 10 second-order section filters. The system supports reloading of filter coefficients during runtime, so that new filters can be implemented on the fly.

Real-time communications between RCG models is provided on several levels: via shared memory for applications running on the same computer; via a PCIe network for computers located within 300m of each other; via a reflected memory network for computers 4km apart. One core on each CPU is reserved for a special model known as the I/O processor (IOP). The IOP handles the interfacing to the I/O hardware modules, and synchronizes the user code execution with the interferometer timing system. The interferometer timing system [54] distributes a 65536 Hz clock signal to all of the IOCs; this clock is synchronized to Global Positioning System time. Timing is also distributed to the front-end computers, to provide time-stamps for real-time communications between RCG models.

The control system also uses the Experimental Physics and Industrial Control System (EPICS) software to provide communications between real-time systems and user interfaces. Though not shown in Figure 21, much of the slow controls hardware (switch settings, slow readbacks, etc.) is managed using an EtherCAT (Ethernet for Control Automation Technology) system.

A custom hierarchical state machine, known as Guardian, manages the global state of the interferometers. Written in Python, Guardian consists of a distributed set of automaton nodes, each handling automation for a distinct sub-domain of the instrument. Each node is loaded with a directed state graph that describes the dynamics of its sub-domain. A master manager node at the top of the hierarchy communicates with multiple sub-manager nodes; these in turn communicate with device level nodes at the bottom of the hierarchy, which directly control the instrument through EPICS. Guardian provides automation of interferometer lock acquisition, as well as the subsequent transitioning to low-noise operation.

## 5. Outlook

Both Advanced LIGO interferometers are currently in the commissioning phase, in which the lock acquisition scheme is implemented and the strain sensitivity is progressively improved. When the sensitivity reaches an astrophysically significant level, the instruments will be operated in observing runs: months- to years-long periods of continuous and coincident data collection. The first Advanced LIGO observing run is planned to begin in 2015, with a duration of 3 months and a BNS detection range of 40—80 Mpc. Subsequent observing runs will have increasing sensitivity and duration. We anticipate a 6 month run in 2016-2017 with a BNS detection range of 80—120 Mpc, and a 9 month run in 2017-2018 at 120—170 Mpc. The full sensitivity corresponding to a 200 Mpc BNS range should be achieved by 2019.

**Acknowledgements**
The authors gratefully acknowledge the support of the United States National Science Foundation for the construction and operation of the LIGO Laboratory and the Science and Technology Facilities Council of the United Kingdom, the Max-Planck-Society, and the State of Niedersachsen/Germany for support of the construction and operation of the GEO600 detector. The authors also gratefully acknowledge the support of the research by these agencies and by the Australian Research Council, the International Science Linkages program of the Commonwealth of Australia, the Council of Scientific and Industrial Research of India, the Istituto Nazionale di Fisica Nucleare of Italy, the Spanish Ministerio de Economía y Competitividad, the Conselleria d'Economia, Hisenda i Innovació of the Govern de les Illes Balears, the Royal Society, the Scottish Funding Council, the Scottish Universities Physics Alliance, The National Aeronautics and Space Administration, OTKA of Hungary, the National Research Foundation of Korea, Industry Canada and the Province of Ontario through the Ministry of Economic Development and Innovation, the National Science and Engineering Research Council Canada, the Carnegie Trust, the Leverhulme Trust, the David and Lucile Packard Foundation, the Research Corporation, and the Alfred P. Sloan Foundation. This article has LIGO document number LIGO-P1400177.